%% file: ChemDFM-R.tex
\documentclass[10pt, letter, onecolumn]{arxiv}

\usepackage{hyperref}
\usepackage{url}
\usepackage[utf8]{inputenc} 
\usepackage[T1]{fontenc}    
\usepackage{booktabs}       
\usepackage{amsfonts}       
\usepackage{nicefrac}       
\usepackage{microtype}      
\usepackage{xcolor}         
\usepackage{amsmath}
\usepackage{multirow}
\usepackage{adjustbox}
\usepackage{lineno}
\usepackage{colortbl}
\usepackage{natbib}

\usepackage{lineno} 
\usepackage{cleveref}
\usepackage{overpic}
\usepackage{wrapfig}
\usepackage{pifont}
\usepackage{array}
\usepackage{amssymb}
\usepackage[textsize=tiny]{todonotes}
\usepackage{caption}
\setuptodonotes{inline}
\todostyle{fixed}{prepend, caption=FIXED, color=green}

\definecolor{darkblue}{rgb}{0, 0, 0.5}
\definecolor{red}{RGB}{255,0,0}
\definecolor{green}{RGB}{0,176,80}
\definecolor{yellow}{RGB}{197,90,17}
\definecolor{grey}{rgb}{0.9,0.9,0.9}
\hypersetup{colorlinks=true, citecolor=darkblue, linkcolor=darkblue, urlcolor=darkblue}

\crefname{section}{§}{§§}
\Crefname{section}{§}{§§}

\title{\Large{ChemDFM-R: A Chemical Reasoning LLM Enhanced with Atomized Chemical Knowledge}}

\author[$\ast$,1]{Zihan Zhao} 
\author[$\ast$,3]{Ziping Wan} 
\author[1,2,3,4,$\dag$]{Lu Chen} 
\author[5]{Xuanze Lin} 
\author[1]{Shiyang Yu}
\author[1]{\\Situo Zhang}
\author[1]{Da Ma}
\author[1]{Zichen Zhu}
\author[1]{Danyang Zhang}
\author[1]{Huayang Wang}
\author[3]{\\Zhongyang Dai}
\author[3]{Liyang Wen} 
\author[3$\dag$]{Bo Chen} 
\author[3]{Xin Chen} 
\author[1,3,4]{Kai Yu}

\affil[1]{\normalsize X-LANCE Lab, School of Computer Science, Shanghai Jiao Tong University, Shanghai, China \authorcr \vspace{0.1cm}}

\affil[2]{\normalsize Shanghai Innovation Institution, Shanghai, China \authorcr \vspace{0.1cm}}

\affil[3]{\normalsize Suzhou Laboratory, Suzhou, China \authorcr \vspace{0.1cm}}

\affil[4]{\normalsize Jiangsu Key Lab of Language Computing, Suzhou, China \authorcr \vspace{0.1cm}}

\affil[5]{\normalsize School of Chemistry and Chemical Engineering, Shanghai Jiao Tong University, Shanghai, China \authorcr \vspace{0.1cm}}

\affil[$\ast$]{\normalsize Equal contributions\hspace{1cm}}
\affil[$\dag$]{\normalsize Corresponding author\authorcr chenlusz@sjtu.edu.cn; chenb@szlab.ac.cn}

\newcommand{\totaltoken}{101 billion}
\newcommand{\papernum}{12 million}
\newcommand{\moleculenum}{30 million}
\newcommand{\reactionnum}{7 million}
\newcommand{\ChemFG}{ChemFG}

\begin{document}

\begin{abstract}
Atomized chemical knowledge, such as functional group information of molecules and reactions, plays a pivotal intermediate role in the reasoning process that connects molecular structures with their properties and reactivities. While large language models (LLMs) have achieved impressive progress, the absence of atomized chemical knowledge results in their superficial understanding of chemistry and limited chemical reasoning capabilities. In this work, to tackle this problem, we develop a Chemical Reasoning LLM, \textbf{ChemDFM-R}. We first construct a comprehensive dataset of atomized chemical knowledge, \textbf{\ChemFG{}}, annotating the presence of functional groups in molecules and the changes of functional groups during chemical reactions, to enhance the model's understanding of the fundamental principles and internal logic of chemistry. Then, we propose a \textbf{mixed-source distillation} method that initializes the model's reasoning capability with limited distilled data, and develop a four-stage training pipeline to equip the model with atomized chemical knowledge and chemical reasoning logic. Experiments on diverse chemical benchmarks demonstrate that \textbf{ChemDFM-R achieves cutting-edge performance while providing interpretable, rationale-driven outputs, surpassing both the general-domain LLMs and domain-specific chemical LLMs}. Moreover, ChemDFM-R achieves comparable or superior performance compared with cutting-edge commercial LLMs, such as o4-mini. Further case studies illustrate how explicit reasoning chains significantly improve the model's reliability, transparency, and practicality in real-world human-AI collaboration scenarios.
\end{abstract}

\maketitle

\section{Introduction}

In the chemical world, functional groups play a pivotal role. As fine-grained atomized chemical knowledge, the presence and transformations of functional groups serve as critical connectors between molecular structures and high-level chemical phenomena, such as reactions and physicochemical properties. For example, the phenol, which consists of a phenyl group bonded to a hydroxy group, is water-soluble, while the toluene, which substitutes the hydrophilic hydroxy group of the phenol with the hydrophobic methyl group, is water-insoluble.
These functional groups constitute essential tools for learning chemical principles, understanding the chemical world, and predicting experimental outcomes. 

Recently, with the remarkable capabilities and performance demonstrated by large language models (LLMs)~\citep{brown2020language, achiam2023gpt, team2023gemini}, great success has been achieved in constructing reasoning LLMs in the general domain~\citep{jaech2024openai, guo2025deepseek, team2025kimi, comanici2025gemini}. Beyond enhancing overall model performance, the reasoning-before-answering pattern directly demonstrates how and why the LLM arrives at the answer, thereby markedly improving the reliability and interpretability of the LLM's response.
However, existing LLMs generally exhibit insufficient mastery of atomized chemical knowledge owing to the shortage of relevant training data.
Even existing chemical LLMs~\citep{zhao2025chemdfm, zhang2024chemllm, zhao2024chemdfmx, zhang2025large, tan2025chemmllm} primarily use literature text and high-level phenomena to enhance their chemical knowledge, while failing to adequately decompose knowledge into underlying mechanisms and atomized chemical knowledge.
As a result, their understanding of chemistry often remains superficial, leading to limited performance on chemical tasks. Moreover, since atomized knowledge serves as a critical intermediate node in chemical reasoning, current general reasoning LLMs struggle to achieve efficient and accurate reasoning for chemical problems. Consequently, their rationales cannot provide sufficient reliability and verifiability.

\begin{figure}[t]
    \centering
    \includegraphics[width=\linewidth]{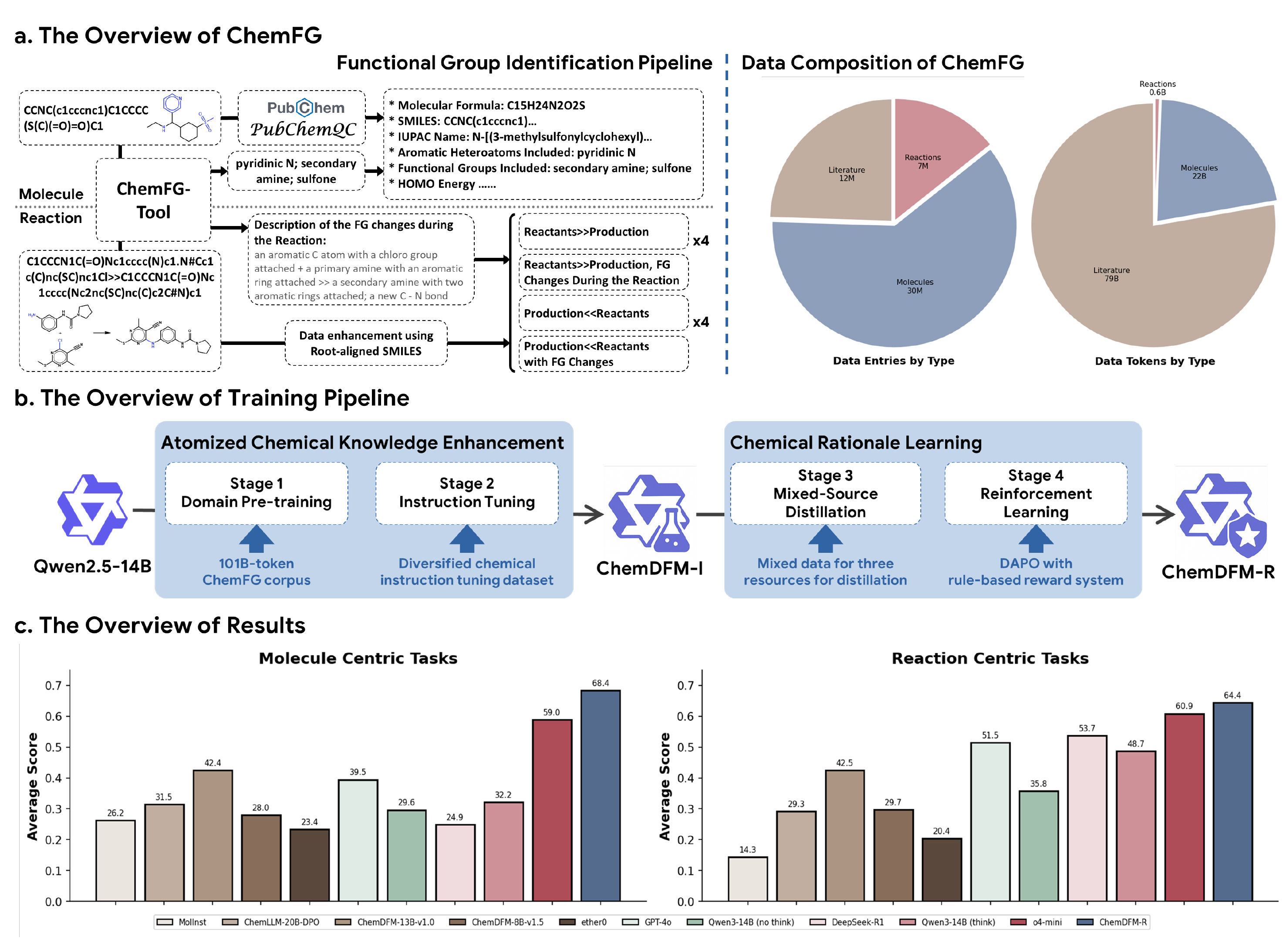} 
    \caption{\textbf{The overview of ChemFG and ChemDFM-R. a. The Overview of ChemFG.} We construct a comprehensive functional group identification toolkit, ChemFG-Tool. The figure demonstrates how the atomized knowledge is annotated and the data composition of the resulting corpus, ChemFG. \textbf{b. The Overview of Training Pipeline.} Our training pipeline is composed of four stages: 1) domain pre-training with the atomized-knowledge-enhanced corpus, ChemFG; 2) instruction tuning on the diversified chemical instruction tuning dataset we construct; 3) Mixed-source distillation that could initialize the reasoning capability with limited distilled data; 4) reinforcement learning with DAPO. \textbf{c. The Overview of Results.} This figure highlights the superior performance of ChemDFM-R on molecule-centric and reaction-centric tasks compared with current chemical LLMs, general non-reasoning LLMs, and general reasoning LLMs, spanning from open-source models to cutting-edge commercial models.}
    \label{fig:overview}
    \vspace{-3mm}
\end{figure}

To better leverage atomized chemical knowledge for building a chemical reasoning LLM, two key challenges must be addressed.
1) Despite the availability of large-scale molecular and reaction databases, comprehensive repositories of atomized chemical knowledge are still missing. Existing datasets predominantly operate at the molecular level, directly presenting the properties or reactivities of molecules. They do not explicitly encode atomized chemical knowledge, such as the presence and transformations of functional groups, to describe these high-level phenomena. In addition, dedicated tools for constructing such atom-level knowledge remain underdeveloped.
2) Distillation data for chemical reasoning based on atomized chemical knowledge are difficult to obtain efficiently. As a common method to initialize the reasoning capability of LLMs, distillation usually involves gathering rationales from advanced reasoning LLMs, such as DeepSeek-R1~\citep{guo2025deepseek} and o3-mini~\citep{openai_o3mini}, and training student models using supervised finetuning. However, owing to the limited understanding of atomized knowledge and chemical logic, even powerful general-domain reasoning models will highly probably fail to generate accurate and in-depth reasoning for chemical problems.

In this work, we focus on addressing the two aforementioned challenges in the chemistry domain and develop a chemical reasoning LLM, \textbf{ChemDFM-R}. Specifically, to tackle the first challenge, we develop a comprehensive functional group identification toolkit and build a large-scale atomized chemical knowledge database with it; to tackle the second challenge, we enhance the generation of distilled data with atomized knowledge and propose a mixed-source distillation method that enables effective distillation training with limited distilled data. The overview of our proposed method is shown in Figure~\ref{fig:overview}.

Firstly, we present \textbf{ChemFG}, a large-scale chemical corpus consisting of both text-based knowledge from chemical literature and atomized chemical knowledge focusing on the presence of functional groups in molecules and the changes of functional groups during reactions. We develop a toolkit, \textbf{ChemFG-Tool}, to comprehensively identify functional groups from molecules and reactions. ChemFG-Tool can effectively identify \textbf{241 types of functional groups} while handling the collision (e.g., the carbonyl group inside a carbamate should not be identified as a ketone or an ordinary ester). The corpus contains over \textbf{\totaltoken{}} tokens from \textbf{\papernum{} literature}, \textbf{\moleculenum{} molecules}, and \textbf{\reactionnum{} reactions}.

Secondly, we develop \textbf{a comprehensive training pipeline} to enhance atomized chemical knowledge and teach models how to reason with it. Specifically, the training pipeline consists of four stages: 1) large-scale domain pre-training on ChemFG, 2) instruction tuning on diverse chemical tasks with diverse instructions, 3) mixed-source distillation that could effectively initialize models' reasoning capability with limited distilled data, and 4) reinforcement learning with Decoupled Clip and Dynamic sAmpling Policy Optimization (DAPO)~\citep{yu2025dapo}.

Thirdly, we introduce \textbf{ChemDFM-R}, an atomized-knowledge-enhanced chemical reasoning LLM, and evaluate it on SciKnowEval~\citep{feng2024sciknoweval} and ChemEval~\citep{huang2024chemeval}.
Extensive experiments demonstrate that ChemDFM-R significantly outperforms both general-domain LLMs and domain-specific chemical LLMs of similar size. On the more difficult benchmark, ChemEval, ChemDFM-R even achieves higher performance (73.8\%) than the cutting-edge commercial LLMs, GPT-4o (63.3\%), DeepSeek-R1 (57.6\%), and o4-mini (67.3\%), highlighting the effectiveness of the atomized chemical knowledge and our proposed training pipeline.
Moreover, we conduct extensive human evaluations to assess the quality of the generated rationales and the potential of ChemDFM-R to facilitate reliable human-AI collaborations. The results show that ChemDFM-R produces more rational and efficient explanations than existing reasoning LLMs. These findings demonstrate that, with the enhancement of atomized chemical knowledge and dedicated training for chemical reasoning, ChemDFM-R can generate high-quality rationales that facilitate deeper understanding and verification of its answers, thereby substantially enhancing its reliability and interpretability.

\section{Results}

\subsection{The Construction of \ChemFG}\label{chemfg}

In the field of chemistry, functional groups serve as the bridge between molecular structures, properties, and reactivities, making them one of the most critical intermediate reasoning steps in chemical reasoning.
However, existing training corpora of LLMs often lack detailed information on molecular functional groups, preventing models from directly and precisely learning this atomized chemical knowledge. Therefore, we collect a functional-group-centered domain pretraining corpus, \ChemFG{}, which consists of data from three sources: literature, molecules, and reactions. The functional groups identification pipeline and basic statistics of \ChemFG{} are shown in Figure~\ref{fig:overview}a with details provided in Section~A.1 of the Supplementary Information.

\paragraph{Functional Group Identification.} Despite the Internet-scale publicly available molecule and reaction corpora, there are no existing databases that describe the correspondence between functional groups and molecules or reactions. To tackle this issue, we develop a functional group identification toolkit, ChemFG-Tool, based on \texttt{thermo} library~\citep{thermoThermoThermodynamics} by extending its embedded SMARTS~\citep{daylightDaylightTheory} list from 83 types of functional groups to 241 and improving its algorithms. With the help of ChemFG-Tool, we annotate the functional groups of all our domain-pretraining {\em molecule} data.
Further details are provided in Section~A.2 of the Supplementary Information.

As for {\em reactions}, we annotate the changes of functional groups during reactions with the following process.
First, with the help of atom mapping annotations provided by the USPTO-FULL dataset, we identify the reaction centers as the atoms that are involved in bond changes during reactions. Based on these reaction centers and our functional group identification toolkit, ChemFG-Tool, we identify the functional groups of the reactants that directly participate in the reaction and those of the product that directly result from the reaction. Finally, the reaction can be described as a functional group transformation, where reacting functional groups are converted into product functional groups. Besides functional groups, there are other structural changes during reactions that are equally important, including ring breaking, ring forming, and bond changes outside functional groups. Therefore, we also construct tools to identify these changes in a similar manner.

\paragraph{Quality Control.} To ensure the annotation quality of functional groups, we hire three graduate-level chemical experts to conduct manual inspections. Firstly, all the experts agree that the extended SMARTS list has already covered the most common functional groups. For molecules, our tool's annotation accuracy of 100 random samples reaches 98\%, with errors primarily due to corner cases such as rare functional groups or complex interactions between functional groups and aromatic rings. For the annotation of reactions, our tool achieves 89\% accuracy when tested with 100 random samples.
The errors mainly arise from invalid reactions or wrong atom mapping annotations. Examples and analyses of the error are demonstrated in Section~A.3 of the Supplementary Information.

\subsection{The Overview of ChemDFM-R}

To develop ChemDFM-R, we built upon Qwen2.5-14B~\citep{yang2024qwen25}, an advanced general-purpose LLM. As outlined in Figure \ref{fig:overview}b, the training pipeline of ChemDFM-R can be divided into two parts: 1) \textbf{Atomized Chemical Knowledge Enhancement}, where the basic general LLM is trained with atomized chemical knowledge; 2) \textbf{Chemical Rationale Learning}, where the model's chemical reasoning capability is enhanced. 

\textbf{Atomized chemical knowledge enhancement} consists of two stages: \emph{Domain Pretraining} and \emph{Instruction Tuning}. During \emph{domain pretraining}, the model is trained on the \ChemFG{} corpus using the next-token-prediction loss, during which the model learns the basic chemical knowledge from the literature and functional-group breakdowns. For \emph{instruction tuning}, we construct a dataset with 16 tasks spanning from name translation and literature question-answering~(QA) to molecular property ordering and reaction completion. Moreover, we diversify the task instructions extensively to enhance the model's generalizability.
Based on Qwen2.5-14B, atomized chemical knowledge enhancement results in an intermediate model named ChemDFM-I.

\textbf{Chemical rationale learning} consists of two stages: \emph{Mixed-Source Distillation} and \emph{Reinforcement Learning}. To efficiently and effectively initialize the model's reasoning ability, we propose a specialized distillation method called \emph{mixed-source distillation}. Specifically, instead of using the distilled chain of thought~(COT) alone during the training, we use 1) the data from the instruction tuning dataset to maintain the chemical knowledge and capabilities; 2) pseudo-reasoning data to enhance the model's functional-group-breakdown capability; and 3) distilled COT to introduce general reasoning patterns and abilities to the model. During \emph{reinforcement learning}, we use the DAPO~\citep{yu2025dapo} algorithm with a rule-based reward system to unify the capabilities learned from the three data resources and further enhance the model's reasoning capability. Based on ChemDFM-I, chemical rationale learning further produces the final model, ChemDFM-R.



\subsection{ChemDFM-R enhances performance on chemical benchmarks}\label{sec:baselines}

We evaluate ChemDFM-R and the baseline models on two of the most popular and comprehensive benchmarks specifically designed for assessing the chemical capabilities of LLMs: SciKnowEval~\citep{feng2024sciknoweval} and ChemEval~\citep{huang2024chemeval}. Given the large number of tasks included in SciKnowEval~(19 tasks) and ChemEval~(35 tasks), to facilitate fair and clear comparison, we categorized the tasks into three groups: text-centric, molecule-centric, and reaction-centric tasks. Details of the task categorization are provided in Section~F of the Supplementary Information.


\begin{figure}
    \centering
    \vspace{-2mm}
    \includegraphics[width=\linewidth]{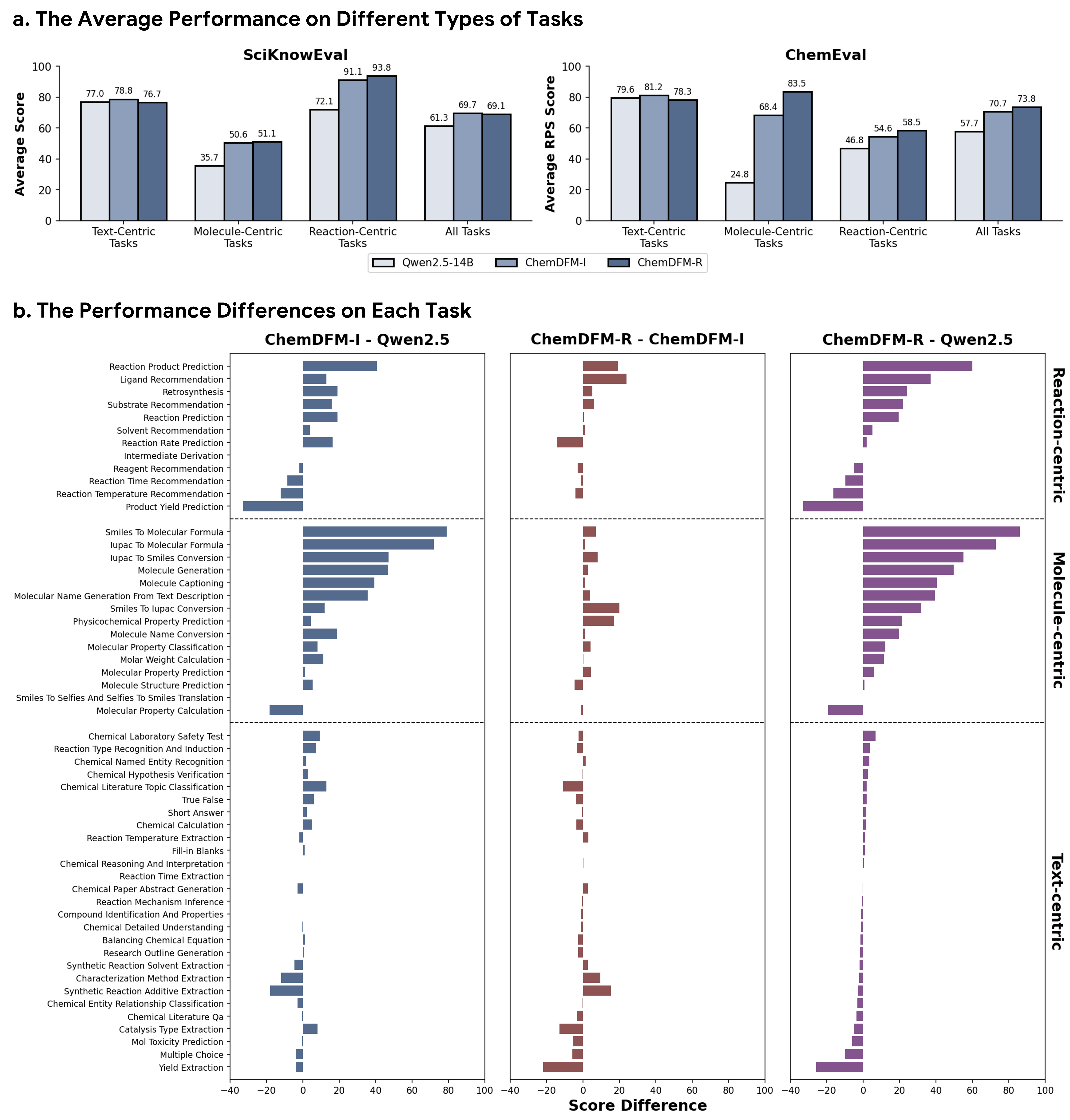} 
    \caption{\textbf{Performance comparison between Qwen2.5-14B-Instruct, ChemDFM-I, and ChemDFM-R.} a. The average performance on different types of tasks of Qwen2.5-14B-Instruct, ChemDFM-I, and ChemDFM-R. b. The performance differences between Qwen-2.5-14B-Instruct, ChemDFM-I, and ChemDFM-R on each individual task of SciKnowEval and ChemEval.}
    \label{fig:compare}
    \vspace{-5mm}
\end{figure}

First, we show the effectiveness of our training pipeline by comparing the performances of ChemDFM-R with those of 1) Qwen2.5-14B-Instruct~\citep{yang2024qwen25}, which is the general-domain instruction tuning model of Qwen2.5-14B, and 2) ChemDFM-I, which incorporates atomized chemical knowledge enhancement but precedes the stage of chemical rationale learning. The quantitative results are illustrated in Figure~\ref{fig:compare}a, while examples of the ChemDFM-R's rationales are analyzed in Section~E of the Supplementary Information.

As showcased in Figure~\ref{fig:compare}a, ChemDFM-R consistently outperforms Qwen2.5-14B-Instruct on both SciKnowEval and ChemEval, demonstrating that our specialization pipeline has successfully improved the model's chemical capabilities. Specifically, the performances on text-centric tasks remain almost intact, while those on molecule-centric and reaction-centric tasks increase significantly. This proves that our method manages to improve the chemical capabilities of LLM while largely maintaining its abilities in understanding natural language. 


Moreover, we also evaluated ChemDFM-I to illustrate the contributions of the different stages in our training pipeline. Results show that \textbf{the atomized chemical knowledge enhancement stage consistently improves performance across all task categories}, while \textbf{the chemical rationale learning stage further strengthens performance on molecule- and reaction-centric tasks}. We attribute this phenomenon to two factors. \textit{First}, the molecule- and reaction-centric tasks typically demand more domain-specific chemical reasoning, such as molecular property prediction or retrosynthesis analysis. In contrast, text-centric tasks rely more on natural language understanding, such as chemical named entity recognition and literature question-answering. As a result, learning chemical reasoning over molecules and reactions provides limited benefit to these text-focused tasks. \textit{Second}, for the sake of answer verifiability, the reinforcement learning tasks do not include purely text-based tasks, which may adversely affect the model’s text reasoning ability. Incorporating text-related tasks into the RL stage through joint training might help preserve performance on text-based tasks.

Furthermore, Figure~\ref{fig:compare}b illustrates the performance changes across individual tasks. The results clearly show that most tasks benefit from our training pipeline, especially the molecule-centric tasks and reaction-centric tasks. Moreover, the two training stages provide complementary gains across different tasks, enabling the final model to achieve superior results on a broader range of tasks. Notably, among the tasks where ChemDFM-R does not surpass Qwen2.5-14B-Instruct, a substantial proportion involves numerical prediction, such as Yield Extraction, Molecular Property Calculation, and Product Yield Prediction. In fact, almost all the molecule-centric and reaction-centric tasks where ChemDFM-R falls short of Qwen2.5-14B-Instruct are those involving numerical reasoning and prediction. This pattern suggests that the numerical calculation and prediction abilities of ChemDFM-R are relatively weak, highlighting a potential direction for further improvements.

\subsection{ChemDFM-R outperforms advanced LLMs}

To further demonstrate the prowess of ChemDFM-R, we compare it with three sets of models: 1) existing chemical LLMs, including MolInst~\citep{fang2023mol}, ChemLLM~\citep{zhang2024chemllm}, ChemDFM~\citep{zhao2025chemdfm}, and ether0~\citep{narayanan2025training}; 2) advanced non reasoning LLMs in the general domain, including GPT-4o~\citep{hurst2024gpt4o} and Qwen3-14B (no think)~\citep{yang2025qwen3}; 3) advanced reasoning LLMs in the general domain, including DeepSeek-R1~\citep{guo2025deepseek}, Qwen3-14B (think)~\citep{yang2025qwen3}, and o4-mini~\citep{openai_o3_o4mini}. The experimental results are illustrated in Figure~\ref{fig:overall}.
For detailed performances of individual tasks, please refer to Section~F of the Supplementary Information.

\begin{figure}
    \centering
    \includegraphics[width=\linewidth]{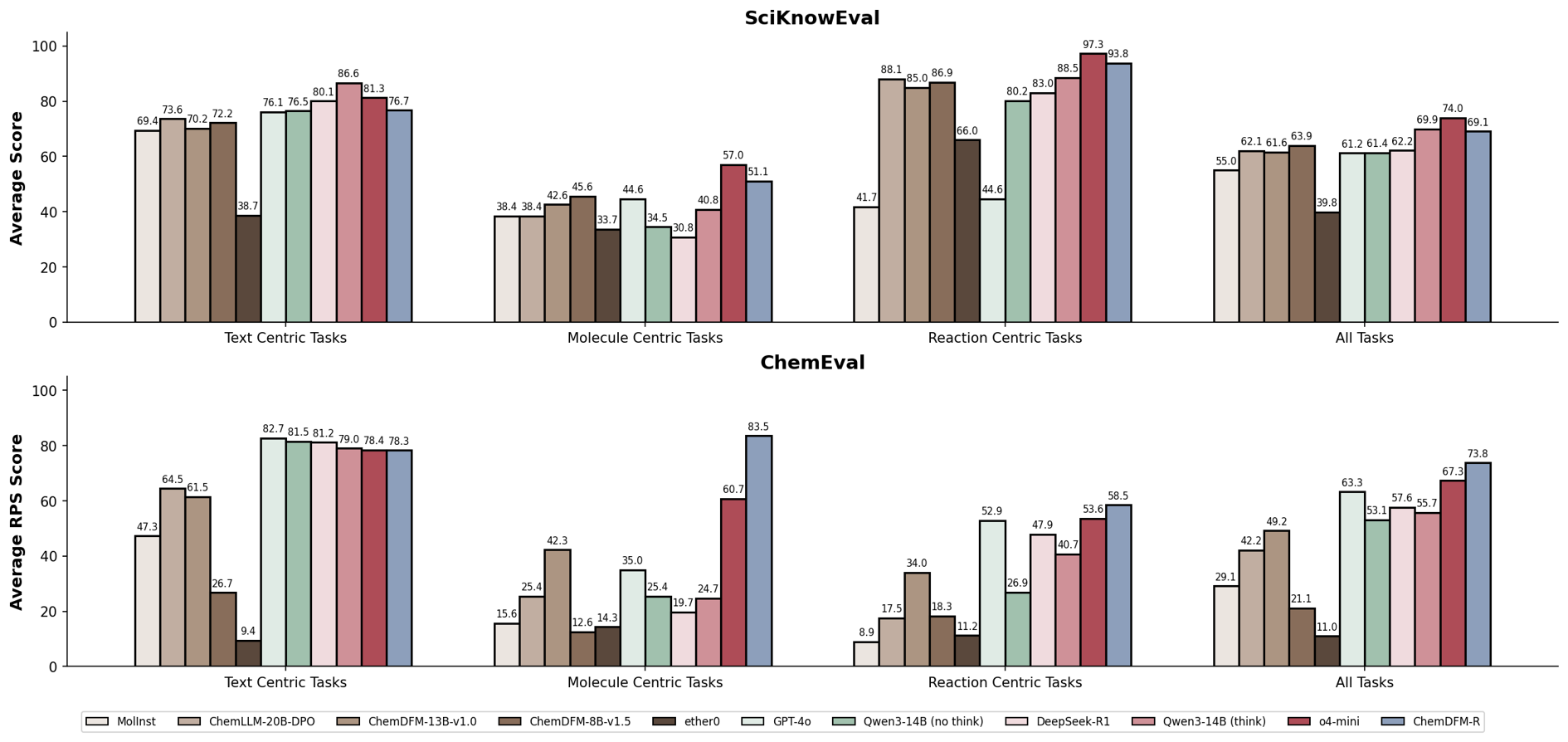}
    \caption{\textbf{Performance comparison of ChemDFM-R and ten baseline LLMs, including chemistry-specific LLMs, general non-reasoning LLMs, and general reasoning LLMs, on SciKnowEval and ChemEval.} We use RPS~\citep{peng2025surveyspeechlargelanguage} to balance the different scales of the scores on different tasks in the ChemEval benchmark.}
    \label{fig:overall}
\end{figure}

As shown in Figure~\ref{fig:overall}, ChemDFM-R significantly outperforms both general-domain LLMs and domain-specific chemical LLMs of similar size, especially on molecule-centric and reaction-centric tasks. In particular, ChemDFM-R achieves an average score of 73.8 on ChemEval, considerably surpassing Qwen3-14B (55.7), the next-generation model in the same series as our base model Qwen2.5-14B. ChemDFM-R also achieves significantly higher performance on molecule-centric and reaction-centric tasks (51.1 \& 93.8) compared to Qwen3-14B (40.8 \& 88.5) on SciKnowEval. These results further validate the effectiveness of the training pipeline of ChemDFM-R and highlight its prowess on chemical tasks. Moreover, existing chemical LLMs perform poorly on these benchmarks, with the best results reaching only 63.9 (on SciKnowEval) and 49.2 (on ChemEval). ChemDFM-R consistently outperforms them across all the task categories on both benchmarks, strongly demonstrating the effectiveness and value of ChemDFM-R and its training pipeline.

When compared to cutting-edge LLMs, ChemDFM-R achieves better performance than GPT-4o and DeepSeek-R1, while demonstrating competitive results relative to o4-mini on SciKnowEval. Considering the tiny size of our model, this result strongly demonstrates the prowess of ChemDFM-R and the effectiveness of our training pipeline.

\subsection{Atomized knowledge is complementary to and more efficient than text-based knowledge}\label{app:fg_abl}

Directly verifying the effect of atomized knowledge enhancement would be extremely costly, since it requires repeatedly performing computationally expensive domain pretraining. Therefore, instead of training a model of the same size as ChemDFM-R on the full dataset for comparison, we used Qwen2.5-1.5B as the base model and conducted the full training pipeline on a 10\% subset of ChemDFM-R's data. By varying the data composition in the subset of the domain-pretraining corpus, we trained different versions of models for comparison while keeping the computational cost manageable. The results are presented in Table~\ref{tab:ablation_fg}.

Compared with the model without any domain pretraining (Row 1), models pretrained on either the text-based-knowledge corpus (Row 2) or the atomized-knowledge corpus (Row 3) show improvements on most tasks. This demonstrates the effectiveness and necessity of domain pretraining for strengthening domain knowledge, and indirectly supports our hypothesis that general domain LLMs generally possess insufficient advanced chemical knowledge. Furthermore, the model pretrained solely on the atomized-knowledge corpus outperforms the model pretrained on the traditional text-based-knowledge corpus on many tasks, despite its corpus being only 20\% of the latter. This provides strong evidence that fine-grained atomized knowledge enables more efficient domain knowledge enhancement. Finally, the model pretrained on the combined text-based- and atomized-knowledge corpora (Row 4) achieves the best overall performance, reflecting the complementary strengths of the two corpora and validating the effectiveness of our atomized-knowledge–enhanced domain pretraining approach.

\begin{table}[]
\small
    \centering
    \caption{Ablation study results on SciKnowEval and ChemEval. DP represents Domain Pretraining, and Know. represent Knowledge. The best performance for each task is indicated using \textbf{boldface}. * We use RPS~\citep{peng2025surveyspeechlargelanguage} to balance the different scales of the scores on different tasks in the ChemEval benchmark.}
    \begin{tabular}{cc|cccc|cccc}
    \toprule
        \multicolumn{2}{c|}{DP Corpus Composition} & \multicolumn{4}{c|}{SciKnowEval} & \multicolumn{4}{c}{ChemEval\textsuperscript{*}} \\
        \cmidrule(lr){1-2} \cmidrule(lr){3-6} \cmidrule(lr){7-10}
        Atomized Know. & Text-based Know. & text & mol. & react. & all & text & mol. & react. & all \\
        \midrule
        \ding{55} & \ding{55} & 66.9 & 30.0 & 34.5 & 49.9 & 44.6 & 25.5 & 25.4 & 34.8 \\
        \ding{55} & \ding{52} & \textbf{67.1} & 30.6 & 36.5 & 50.4 & 43.6 & 25.6 & 28.5 & 35.2 \\
        \ding{52} & \ding{55} & 65.7 & \textbf{31.8} & \textbf{37.8} & 50.3 & 51.9 & 25.9 & \textbf{33.8} & 40.8 \\
        \ding{52} & \ding{52} & 66.2 & 31.7 & 37.4 & \textbf{50.5} & \textbf{53.7} & \textbf{26.9} & 31.0 & \textbf{41.1} \\
        \bottomrule
    \end{tabular}
    \label{tab:ablation_fg}
\end{table}

\subsection{Mixed-source distillation successfully initializes the model's reasoning capability}\label{ablation}

To validate the effectiveness of our newly designed mixed-source distillation method, we conduct an ablation study by gradually simplifying the composition of the distillation dataset. The results are shown in Table~\ref{tab:ablation}. The results prove that the traditional distillation method (Row 2) struggles to achieve positive impacts on performance in the chemical domain. It even underperforms the ``zero'' method (Row 1) proposed by Deepseek-R1, where there is no distillation stage before reinforcement learning. With the help of data sampled from the instruction-tuning dataset to maintain chemical capabilities and knowledge (Row 3), the model's performance gets boosted significantly. Moreover, the pseudo-reasoning data further help the model to achieve higher performance (Row 4, which corresponds to the final setting of our proposed mixed-source distillation method) and achieve better performance than the model from the ``Zero'' method training. These results strongly demonstrate the effectiveness and necessity of our design of mixed-source distillation.

\begin{table}[t]
    \centering
    \caption{Ablation study results on SciKnowEval and ChemEval. ``No Thinking'' denotes the data source of the instruction-tuning dataset, ``Pseudo'' denotes the pseudo-reasoning data source, and ``Distilled'' denotes the teacher’s rationale data source. The best performance for each task is indicated using \textbf{boldface}. * We use RPS~\citep{peng2025surveyspeechlargelanguage} to balance the different scales of the scores on different tasks in the ChemEval benchmark.}
    \label{tab:ablation}
    \begin{tabular}{ccc|cccc|cccc}
    \toprule
    \multicolumn{3}{c|}{Data Source for Distillation} & \multicolumn{4}{c|}{SciKnowEval} & \multicolumn{4}{c}{ChemEval\textsuperscript{*}} \\ 
    \cmidrule(lr){1-3} \cmidrule(lr){4-7} \cmidrule(lr){8-11}
    No Thinking & Pseudo & Distilled & text & mol. & react. & all & text & mol. & react. & all \\
     \midrule
     \ding{55} & \ding{55} & \ding{55} & \textbf{77.7} & 49.9 & 94.2 & 69.2 & 76.4 & 70.9 & 59.9 & 70.4 \\
     \ding{55} & \ding{55} & \ding{52} & 75.5 & 49.6 & 87.9 & 67.2 & 72.0 & 68.8 & \textbf{63.6} & 68.9 \\
     \ding{52} & \ding{55} & \ding{52} & 77.4 & 50.6 & 92.4 & 69.1 & 78.0 & 83.1 & 61.8 & 74.5 \\
     \ding{52} & \ding{52} & \ding{52} & 76.8 & \textbf{52.0} & \textbf{94.5} & \textbf{69.5} & \textbf{80.3} & \textbf{84.5} & {61.3} & \textbf{75.8} \\
     \bottomrule
    \end{tabular}
\end{table}

\subsection{ChemDFM-R generates more accurate, reliable, and user-friendly rationales}

To assess the quality of rationales under practical situations, we constructed ten graduate-level questions based on recent publications from several influential chemistry journals. The questions cover different major subfields of chemistry, including organic chemistry, inorganic chemistry, materials chemistry, analytical chemistry, and polymer chemistry. Then, different reasoning LLMs are leveraged to solve these questions through human-AI interactions. Five graduate-level chemistry experts were hired to evaluate these interactions across five dimensions, each of which is scored on a 5-point scale:

\begin{itemize}
    \item Chemical Correctness: The correctness of chemical knowledge and logic demonstrated throughout the reasoning process.
    \item Answer Accuracy: The accuracy of the final answer given by the model.
    \item Analytical Coverage: The extent to which different plausible possibilities are explored during reasoning.
    \item Reasoning Coherence: Whether the reasoning remains focused, coherent, and aligned with the problem.
    \item Effective Information Density: The density of useful information in the reasoning chain, reflecting the friendliness and efficiency of interaction.
\end{itemize}

The evaluation results are demonstrated in Figure~\ref{fig:human}, while the questions are listed in Section~G of the Supplementary Information. From these results, we draw three main conclusions:

First, our model outperforms all baselines, including DeepSeek-R1, in both chemical correctness and answer accuracy, indicating that it possesses a stronger grasp of chemical knowledge. By carefully analyzing and comparing the evaluation results across question types, we observe that our model shows a clearer advantage on SMILES-related tasks, demonstrating a more precise understanding of molecular structures.

Second, our model achieves significantly higher scores in effective information density. Unlike models such as Qwen3-14B and DeepSeek-R1, which often generate extremely lengthy reasoning chains exceeding 1,000 tokens, our model produces concise yet informative reasoning. This considerably reduces the burden on human users when verifying correctness or identifying errors.

Finally, our model performs slightly lower than Qwen3-14B and DeepSeek-R1 on analytical coverage. This is partially because their extremely lengthy reasoning chains allow them to enumerate many possible considerations. It is also true that our model commonly focuses on one or a few major possibilities rather than adequately enumerating all potential factors. This highlights a productive direction for future enhancement of our model’s analytical breadth.

\begin{figure}
    \centering
    \includegraphics[width=\linewidth]{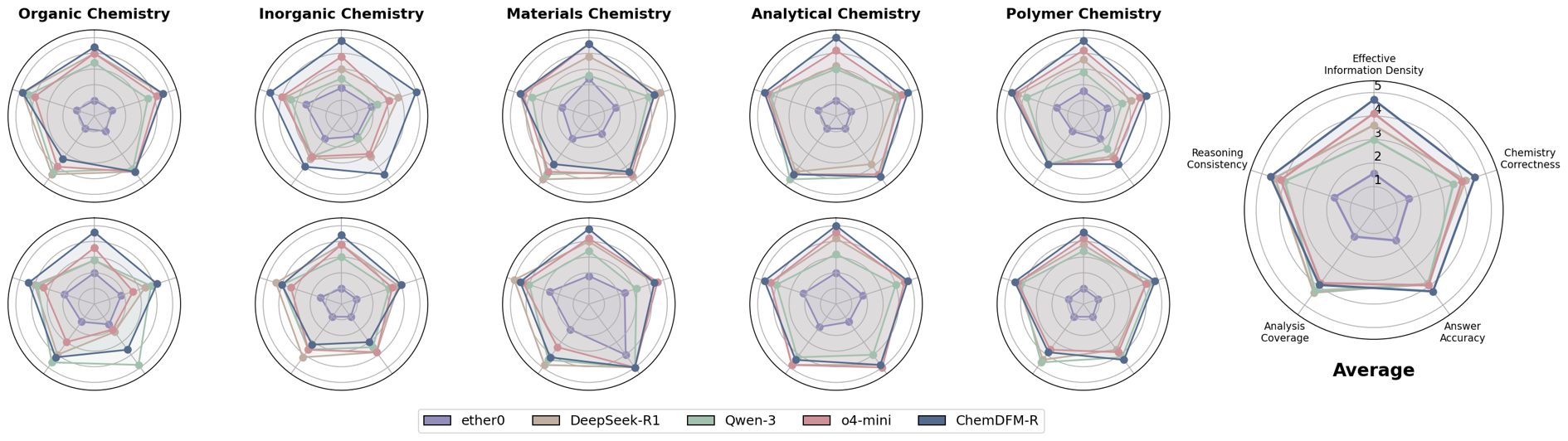}
    \caption{\textbf{Human assessment of the quality of rationales from different reasoning LLMs.} Each small radar chart on the left corresponds to the scores of a question during the human assessment, and the large radar chart on the right is the average scores of the models on all ten questions. Please refer to Section~G of the Supplementary Information for the detailed questions used in this assessment. As the rationales of o4-mini are not accessible, we only use the answer provided by o4-mini. All metrics are higher-is-better.}
    \label{fig:human}
\end{figure}
 
\begin{figure}
    \centering
    \vspace{-3mm}
    \includegraphics[width=\linewidth]{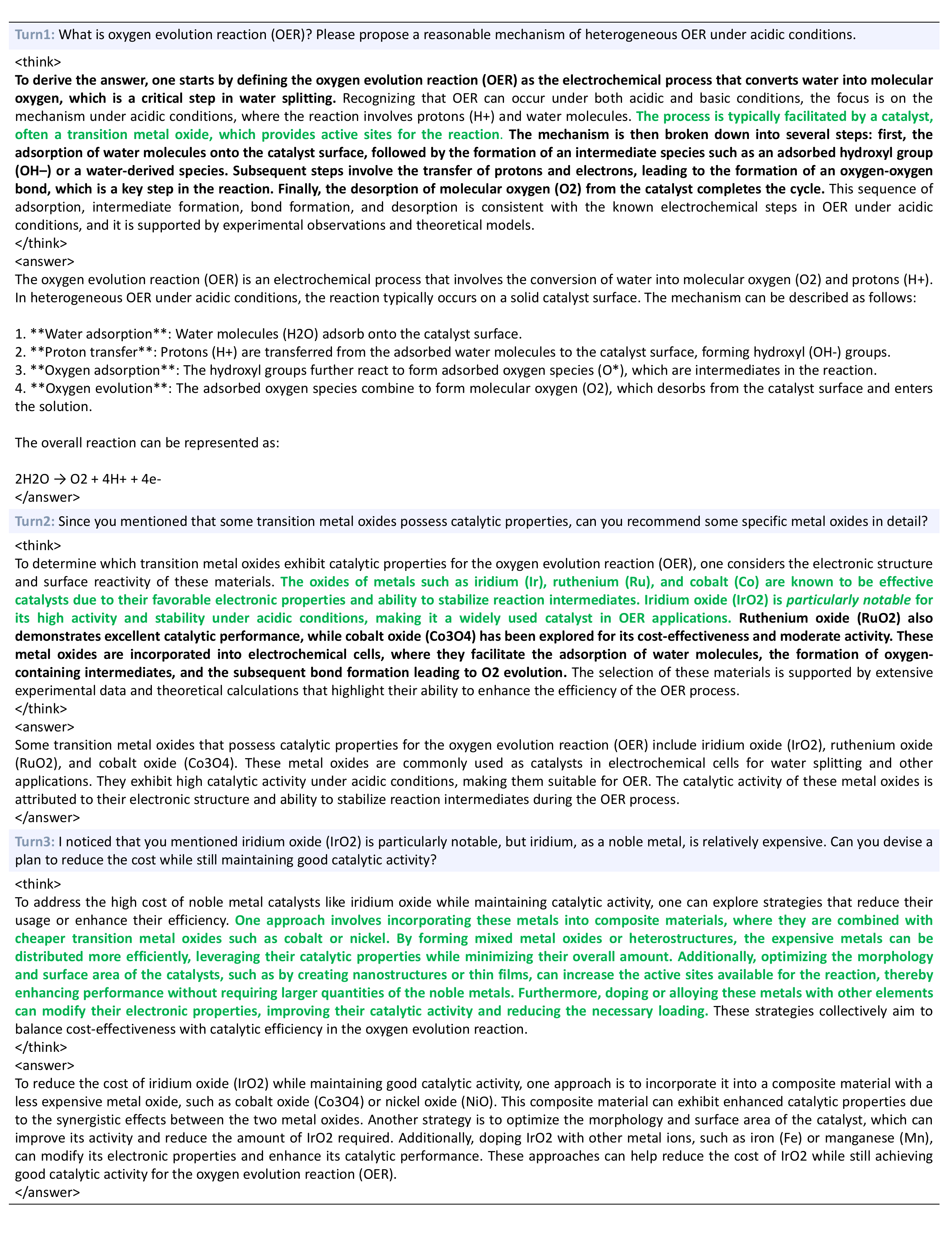} 
    \vspace{-7mm}
    \caption{An example of reliable human-AI collaboration using ChemDFM-R. This conversation is inspired by Li et. al. \citep{li2025achieving}. This example illustrates a process of research inspiration seeking with the help of ChemDFM-R.}
    \label{fig:collab}
    \vspace{-5mm}
\end{figure}

\subsection{ChemDFM-R facilitates more reliable human-AI collaborations}

Another important advantage of ChemDFM-R's reasoning capability is that it allows humans to verify the correctness of answers, identify and correct errors, and discover new insights or perspectives. This enables more practical, reliable, and flexible human-AI collaboration. We then demonstrate the prowess of ChemDFM-R in enabling reliable human–AI collaboration through illustrative examples. Specifically, an example is showcased in Figure~\ref{fig:collab}, while more examples are illustrated in Section~H of the Supplementary Information.

To make our example more realistic, we design our conversation topic according to a newly published chemical research paper \citep{li2025achieving}. It is worth noticing that this paper was published in 2025, so it is impossible for ChemDFM-R to encounter this paper during either the pretraining of the base model, Qwen2.5-14B, or our domain-pretraining process. To demonstrate the potential of ChemDFM-R, we assume the role of a researcher seeking a new research inspiration and engage the model in a dialogue about the subfield to which this paper belongs.

First, as a researcher seeking inspiration, we ask ChemDFM-R to introduce the oxygen evolution reaction~(OER) briefly, including the definition and mechanism of it. Although the answer itself is a relatively rigid response to the question, the model’s rationale reveals interesting insights. In its rationale, ChemDFM-R mentions that OER requires a transition metal oxide catalyst, which often presents valuable opportunities for further research. Therefore, we further ask it to give more detailed recommendations for the catalysts. As a response, ChemDFM-R proposes multiple different oxides, which is a wide range for us to dig into. However, in its rationale, ChemDFM-R itself says that ``iridium oxide (IrO2) is \textit{particularly notable} for its ...'', which is very inspiring. Since it is well known that iridium-based compounds are often very expensive, a natural follow-up question arises: how can we optimize this catalyst to reduce its cost while maintaining its catalytic performance? Surprisingly, ChemDFM-R manages to propose the initial ideas that align closely with the ideas presented in Li et. al. \citep{li2025achieving}, ``forming mixed metal oxides or heterostructures'' and ``optimizing the morphology and surface area of the catalysts''. At this point, a broad research direction has taken shape.

It is worth noticing that nearly all the inspirations are drawn from the rationale of ChemDFM-R, which demonstrates the significance and value of ChemDFM-R’s ability to generate reasoning. This example shows that, with the enhanced chemical knowledge and strong chemical reasoning capabilities, ChemDFM-R has the potential to facilitate reliable human-AI collaboration, thereby advancing AI-driven research and applications. More examples involving error correction and answer improvement with the help of rationales are demonstrated in Section~H of the Supplementary Information.



\section{Discussion}

In this paper, we identify a key factor limiting the performance of existing LLMs and present ChemDFM-R, a new chemical reasoning LLM that achieves advanced performance. Specifically, we argue that the insufficient mastery of atomized chemical knowledge is a key factor underlying the limited performance of current LLMs, both general domain LLMs and current chemical LLMs, on chemical tasks. To tackle this problem, we develop a comprehensive toolkit, ChemFG-Tool, for functional group identification and construct a large-scale atomized-knowledge–enhanced chemical corpus, ChemFG. Leveraging ChemFG-Tool and ChemFG, we develop a comprehensive training pipeline that 1) enhances the model's understanding of atomized chemical knowledge with domain pre-training and instruction tuning, and 2) teaches the model the key reasoning logic and method in chemistry with mixed-source distillation and reinforcement learning.

The ChemDFM-R model achieves advanced results across multiple tasks in our comprehensive evaluations of different benchmarks. It consistently outperforms all models of comparable size and, despite having only 14B parameters, achieves performance that is competitive with or even surpasses state-of-the-art proprietary models. These results strongly illustrate both the effectiveness of our training pipeline and the prowess of the resulting model, ChemDFM-R.

To better validate the effectiveness and necessity of the key designs of our training pipeline, we 1) compare ChemDFM-R's performance with ChemDFM-I and Qwen2.5-14B-Instruct, which directly demonstrate the impact of our training process, and 2) design comprehensive ablation studies, validating the necessity of our design. From these evaluations, two main conclusions are drawn. Firstly, atomized knowledge enhancement could effectively and efficiently improve performance across all task categories. It achieves similar or even larger performance gains with only 20\% of the tokens of the text-based knowledge. The best performance is achieved by combining atomized knowledge and text-based knowledge, illustrating the complementary strength of the two. Secondly, chemical rationale learning strengthens the performance on molecule- and reaction-centric tasks. We also validate the effectiveness and necessity of the mixed-source distillation design for initializing models' reasoning capabilities when the distilled data is limited. Although the chemical rationale learning effectively endows the model with reasoning capability while also improving its overall performance, it weakens the model's general language abilities, especially numerical prediction abilities. This is mainly because our RL stage relies on rule-based methods to calculate rewards, which restricts the types of tasks we can use. Designing reward mechanisms better suited to chemical contexts and allowing a broader range of RL tasks will be an important direction for the improvement of ChemDFM-R.

Besides the quantitative evaluation, we also conduct human assessments and case studies to 1) validate that ChemDFM-R could provide more precise and focused chemical rationales, and 2) illustrate the value of ChemDFM-R's rationales in reliable human-AI collaborations under real-world scenarios. The results show that ChemDFM-R's rationales are not only more precise and accurate, but also more concise and informative. This eliminates the need for researchers to go through reasoning chains containing thousands of tokens to locate valid reasoning and supporting evidence. Consequently, ChemDFM-R provides more reliable and verifiable answers, enhancing both usability and efficiency. We demonstrate ChemDFM-R's potential in real-world human–AI collaboration by case studies, showcasing the critical role of its rationales in insight discovery, answer verification, and error detection and correction. However, although our model produces precise and informative reasoning chains, this conciseness can compromise analytical coverage. Achieving a better balance between maintaining concise reasoning chains and providing sufficiently diverse analyses could further enhance ChemDFM-R's practical value in interactive settings.

Looking ahead, this work reveals a new perspective and methodology for training domain-specific LLMs. Traditionally, efforts to improve the performance of domain-specific LLMs have focused on increasing the size of the training corpus or enhancing the quality of the knowledge used during training. In this work, we point out that the learning efficiency of domain models varies depending on the form and granularity of the provided knowledge. Consequently, improving the efficiency of knowledge acquisition represents an equally important avenue for enhancing domain models' performance. In this study, we achieve this by modifying the format of the knowledge-bearing corpus and incorporating finer-grained atomized domain knowledge, which significantly improves the model’s learning efficiency and leads to stronger overall performance. Similar strategies can be applied to other domains. Identifying the optimal form and granularity of domain knowledge for LLMs can yield improvements comparable to or even more efficient than increasing corpus size. Such approaches can either complement larger corpora or serve as effective alternatives when computational resources are limited. These insights gained from this work may inform the specialization training of LLMs across diverse domains, providing alternative methods and tools for developing stronger domain-specific LLMs.

\begin{figure}[t]
    \centering
    \includegraphics[width=0.8\linewidth]{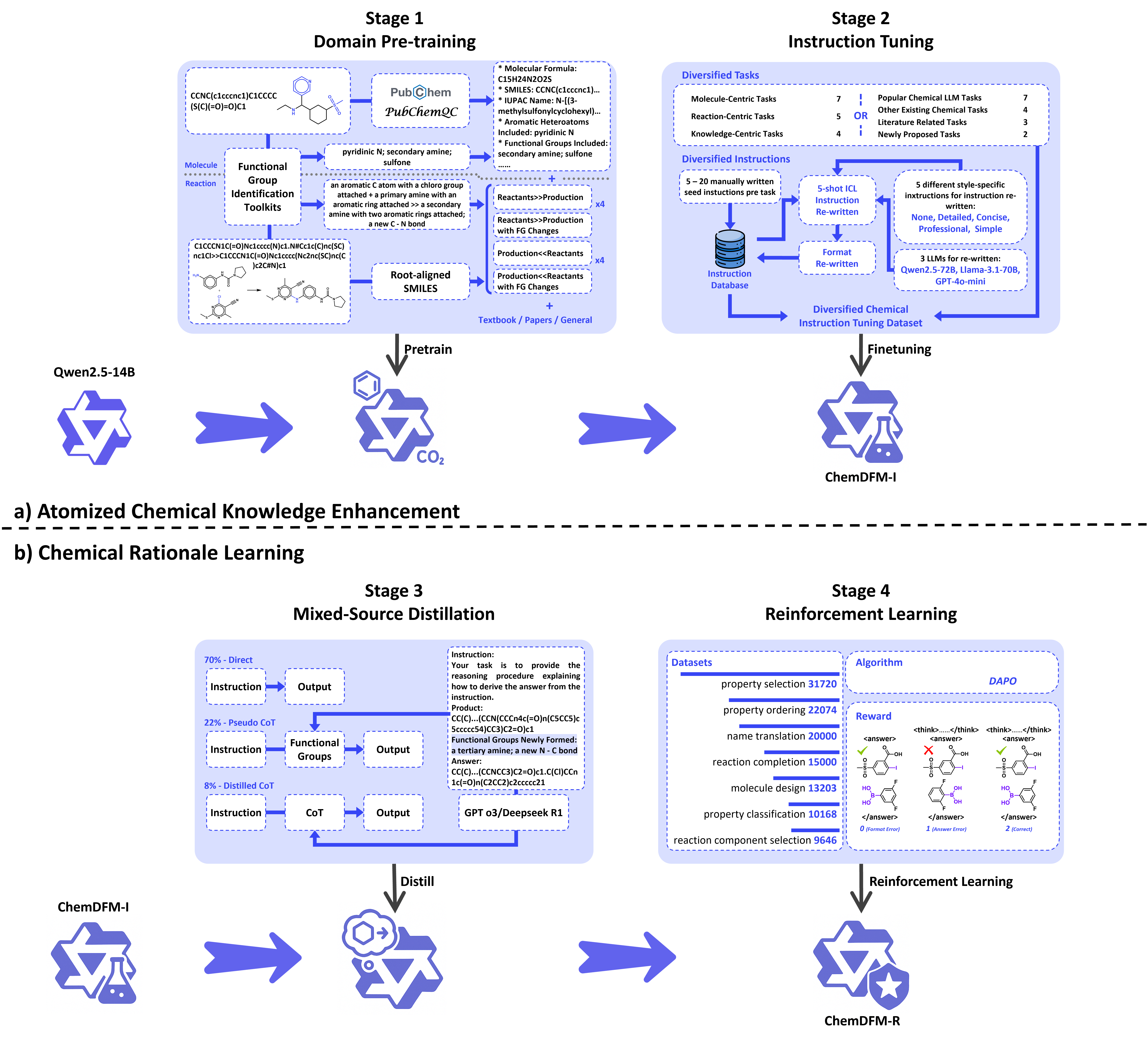} 
    \caption{\textbf{The details of ChemDFM-R's training pipeline.} The four-stage training pipeline can be divided into two parts: \textbf{a. Atomized Chemical Knowledge Enhancement} where atomized chemical knowledge is introduced into the model, including 1) domain pre-training with the atomized-knowledge-enhanced corpus, ChemFG; 2) instruction tuning on the diversified chemical instruction tuning dataset we construct, which contains diversified chemical tasks and diversified chemical instructions; \textbf{b. Chemical Rationale Learning} where the model learns to reason in chemistry with atomized knowledge, including 3) Mixed-source distillation that could initialize the reasoning capability with limited distilled data; 4) reinforcement learning using DAPO with the rule-based reward system.}
    \label{fig:train_pipe}
    \vspace{-3mm}
\end{figure}

\section{Methods}

We present the major details of our four-stage training pipeline for ChemDFM-R~(Figure~\ref{fig:train_pipe}), while more details about the data-leakage prevention, hyperparameter choosing, and system prompts are introduced in Section~D of the Supplementary Information.

\subsection{Atomized Chemical Knowledge Enhancement}\label{dpit}

In this stage, our model mainly learns the atomized chemical knowledge to prepare itself with ``ingredients'' to ``cook'' the chemical rationales. Specifically, we achieve that through domain pretraining and instruction tuning.

\paragraph{Domain Pretraining.} In domain pretraining, we leverage the \totaltoken-token \ChemFG{} corpus introduced in Section~\ref{chemfg} to familiarize our model with the knowledge related to functional groups. We train our model from one of the most advanced general LLMs, Qwen2.5-14B~\citep{yang2024qwen25}. Considering that general knowledge is also vital for Chemical LLMs, we also incorporate a substantial amount of general-domain pretraining data into our domain pretraining corpus to ensure that the model retains its general capabilities as much as possible.

\paragraph{Instruction Tuning.} The primary goal of instruction tuning is to teach the model how to analyze the purpose and requirements of a given task and make proper use of the knowledge learned in the pretraining phase. However, existing instruction tuning datasets in the field of chemistry are typically derived from well-studied chemistry tasks and suffer from a severe lack of diversity of both task varieties and instruction expressions.
Therefore, we construct a new instruction tuning dataset for ChemDFM-R based on the instruction tuning dataset of ChemDFM~\citep{zhao2025chemdfm}. To improve the overall task and instruction diversity, we introduce numerous new chemistry-related tasks, such as scientific paper QA, chemical property ordering, and reaction step prediction, and perform instruction-rewriting to achieve an average instruction-entry ratio of 1:50. For detailed information on the construction and composition of the instruction tuning dataset, please refer to Section~B of the Supplementary Information.
To maintain the general capabilities of our model, we mixed our chemical instruction tuning data with general instruction tuning data in a 1:2 ratio. The Qwen2.5-14B model is finetuned for 2 epochs on this mixed dataset after domain pretraining, resulting in the ChemDFM-I model.

\subsection{Chemical Rationale Learning}\label{drl}

\begin{figure}[t]
    \centering
    \vspace{-2mm}
    \includegraphics[width=\linewidth]{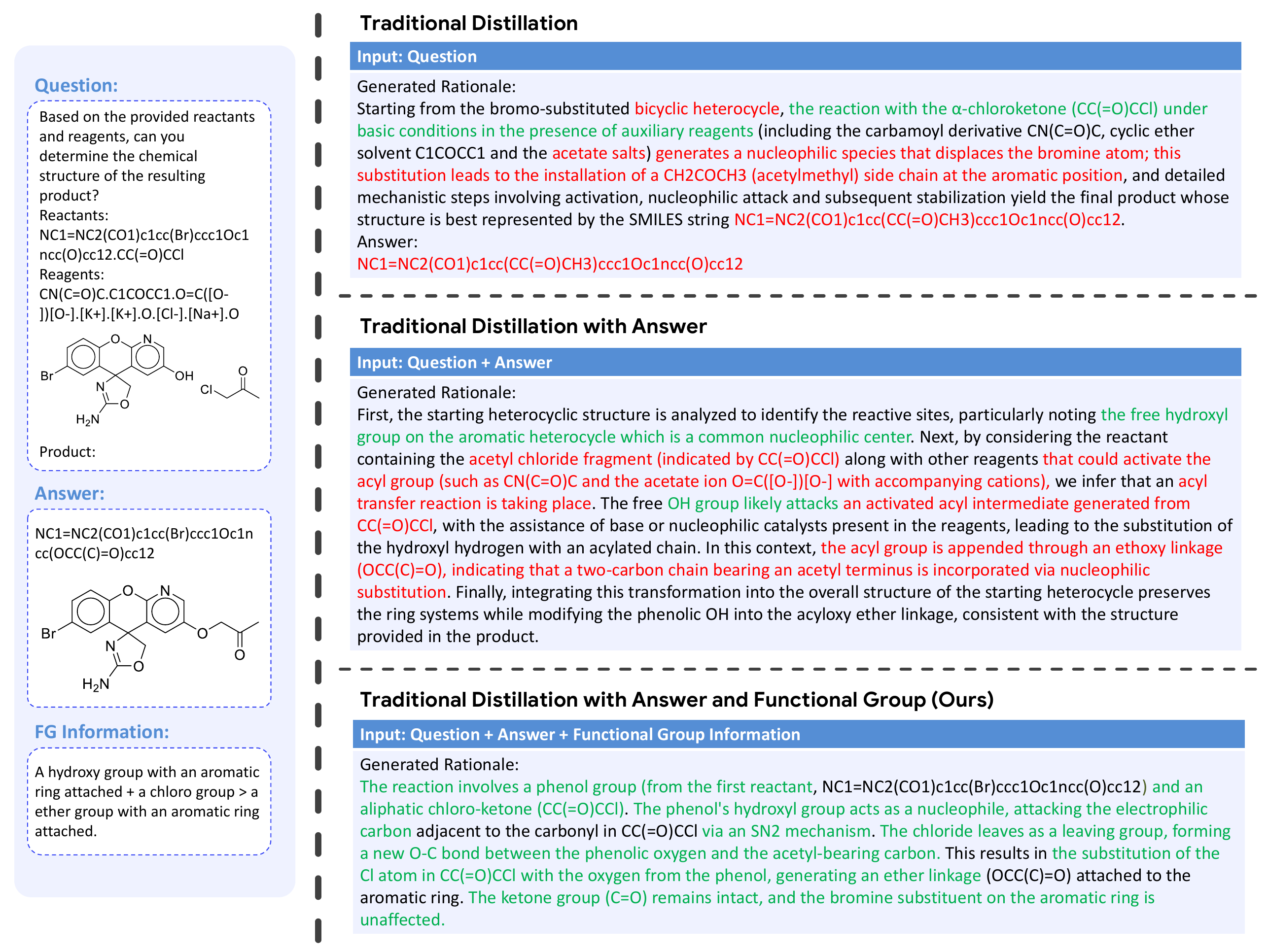} 
    \caption{Comparison of rationales generated by o3-mini with and without additional input information. We mark the correct analyses in the rationale as \textcolor{green}{green}, while the wrong ones as \textcolor{red}{red}. For more examples and detailed analyses, please refer to Section~C.1 of the Supplementary Information.}
    \label{fig:dist_ration_comp}
    \vspace{-5mm}
\end{figure}


The primary goal of this stage is to teach the model how to reason with the atomized knowledge it has acquired. Chemical reasoning requires a deep understanding of chemical principles and logic, as well as the capability to apply them for analysis. These capabilities can not be learned or induced from general-domain reasoning training. Therefore, we propose a chemical rationale learning pipeline to specifically enhance the chemical reasoning capabilities of LLMs based on distillation and reinforcement learning.

\paragraph{Mixed-Source Distillation.} We leverage distillation to prevent the early unstable cold start phase of reinforcement learning. It could illustrate the reasoning patterns to the model and build up its basic reasoning capabilities. 


Specifically, the entries in the distillation dataset come from three sources, each of which corresponds to part of the abilities required for chemical reasoning: 1) the instruction tuning dataset of ChemDFM-R ($\sim$70\%) to maintain basic chemical knowledge and prevent catastrophic forgetting; 2) pseudo-reasoning data describing the functional groups of involved molecules or reactions ($\sim$22\%) and highlighting vital intermediate reasoning steps and functional group analyses; 3) teachers' rationales from DeepSeek-R1 and o3-mini ($\sim$8\%) which introduce general reasoning patterns to the model and initiate its reasoning capabilities.

To improve the quality and efficiency of the teacher's rationales in the chemical domain, instead of asking teacher models to generate rationales from scratch, we provide them with rich additional information. Specifically, the teacher models are provided with the question instruction, the ground truth answer, and the functional group information of the molecules and reactions in the question. Comparison of the rationales generated by DeepSeek-R1 with and without the additional information is illustrated in Figure~\ref{fig:dist_ration_comp}. The rationales generated with full additional information are significantly more valid and in-depth than the other two. More examples with detailed analysis of the rationale generation are given in Section~C.1 of the Supplementary Information.  Moreover, to comprehensively assess the quality of our distilled rationales, we sampled a small subset of them and conducted a quantitative human assessment. The results are illustrated in Section~C.2 of the Supplementary Information.

Similar to instruction tuning, we mix our mixed-source distillation dataset with general data in a 1:2 ratio. The general data are also sampled from multiple sources, where $\sim$92\% of the entries are sampled from the general data for instruction tuning of ChemDFM-R and $\sim$8\% are from AM-Deepseek-R1-Distill-1.4M~\citep{zhao20251}. The ChemDFM-I model is finetuned for one epoch on this mixed dataset.

\paragraph{Reinforcement Learning.} After distillation, reinforcement learning~(RL) is leveraged to further enhance the reasoning capabilities of our model. To construct the reinforcement learning dataset, a subset of tasks that have verifiable answers is selected from our instruction-tuning dataset. The composition of the RL dataset is illustrated in Figure~\ref{fig:train_pipe}. We sampled data from the same sources used to build the instruction-tuning dataset, while minimizing overlap between the specific instances contained in the two datasets. These tasks are mixed together and uniformly sampled during training.

We follow the method recommended by DeepSeek-R1~\citep{guo2025deepseek}, where the reward system consists of format rewards $R_{format}$ and correctness rewards $R_{correct}$. The format rewards evaluate whether a response strictly follows the reasoning format, which is \texttt{``<think>...</think>\textbackslash n<answer>...</answer>''}. The correctness rewards evaluate whether a response is correct by comparing the answer between \texttt{<answer>} and \texttt{</answer>} with the ground truth answer. Specifically, considering the redundancy of the casual SMILES notations (one molecule could correspond to multiple SMILES), we first canonicalize all the SMILES in the response before calculating the accuracy rewards. Formally, for a given instruction and the model's output $o$ and ground truth answer $a$, we define the reward as
$$
R(o, a) = \lambda R_{format}(o) + R_{correct}(o, a),
$$
where $\lambda$ is a coefficient that weights the two parts of the reward and is set to 1 during our training.

We use the Decoupled Clip and Dynamic sAmpling Policy Optimization~(DAPO)~\citep{yu2025dapo} algorithm for our reinforcement learning. For each question $q$ with ground truth answer $a$ from the training dataset $\mathcal{D}$, DAPO algorithm first samples a group of $G$ outputs $\{o_i\}_{i=1}^G$ from the model and then trains the policy model $\pi$ by optimizing the following objective:
\begin{align*}
    \mathcal{J}_{\text{DAPO}}(\theta) 
    &= \mathbb{E}_{(q,a)\sim \mathcal{D}, \{o_i\}_{i=1}^G\sim \pi_{\text{old}}(\cdot|q)} \left[
        \frac{1}{\sum_{i=1}^{G} |o_i|} \sum_{i=1}^{G} \sum_{t=1}^{|o_i|} 
        \min\left(
            r_{i,t}(\theta) \hat{A}_{i,t},\: c\cdot \hat{A}_{i,t}
        \right) 
    \right]  \\
    c &= \text{clip}\left(r_{i,t}(\theta), 1 - \epsilon_{\text{low}}, 1 + \epsilon_{\text{high}} \right) \\
    \text{s.t.} \quad & 0 < \left| \left\{ o_i \mid \text{is\_equivalent}(a, o_i) \right\} \right| < G,\nonumber
\end{align*}
where $\epsilon_{low}$ and $\epsilon_{high}$ denote the clipping range for the importance sampling ratio $r_{i,t}(\theta)$, $\hat{A}_{i,t}$ denote the advantage for the $i$-th response, and $\text{is\_equivalent}(a, o_i)$ measures whether the predicted answer in output $o_i$ matches the ground truth answer $a$. Formally, $r_{i,t}(\theta)$ and $\hat{A}_{i,t}$ are calculated as follows:
\begin{align*}
 r_{i,t}(\theta) = \frac{\pi_\theta(o_{i,t} \mid q, o_{i, <t})}{\pi_{\text{old}}(o_{i,t} \mid q, o_{i, <t})}, \hat{A}_{i,t} = \frac{R_i - \text{mean}(\{R_i\}_{i=1}^G)}{\text{std}(\{R_i\}_{i=1}^G)},
\end{align*}
where $R_i$ denotes the reward for the $i$-th output $o_i$ from $\{o_i\}_{i=1}^G$.

\section{Materials Availability}
The parameters of the model generated in this study, ChemDFM-R, have been deposited in Huggingface: \url{https://huggingface.co/OpenDFM/ChemDFM-R-14B}.

\section{Data Availability}
The functional group identification toolkit, ChemFG-Tool, used during the construction of ChemFG is available at GitHub: \url{https://github.com/OpenDFM/ChemFG-Tool}. The evaluation benchmarks are obtained from the original benchmark authors.

\subsubsection*{Acknowledgments}
This work was supported by the National Science and Technology Major Project (2023ZD0120703), the China NSFC Projects (62576212, 92370206, U23B2057, 62120106006), and the Shanghai Municipal Science and Technology Project (25X010202846).

\subsubsection*{Author Contributions}
Conceptualization, Z. Zhao and L.C.; methodology, Z. Zhao, B.C., and L.C.; software, Z. Zhao, B.C., and D.M.; validation, Z.W., X.L., and S.Y.; data curation, Z. Zhao, B.C., Z.W., S.Z., H.W., Z.D., and W.L.; writing – original draft, Z. Zhao, B.C., Z. Zhu, and D.Z; writing – review \& editing, Z. Zhao, L.C., and D.Z.; supervision, L.C., X.C, and K.Y.

\subsubsection*{Competing Interests}
The authors declare no competing interests.

\bibliography{custom}
\bibliographystyle{unsrt}



\input{appendix}

\end{document}

%% file: appendix.tex
\appendix

\section{Details about \ChemFG}\label{app:chemfg}

\subsection{Raw Data Collection}\label{statistics}

\paragraph{Literature.} Literature, including papers and textbooks, contains not only the widely accepted chemical knowledge and principles, but also the cutting-edge research in the field of chemistry. Therefore, to take full advantage of chemical literature, we collected over \papernum{} literature from the open Internet dated prior to January 2022. After further cleaning and deduplication, 79B tokens are obtained from it.

\paragraph{Molecules.} Molecules are the fundamental participants in various chemical processes. Therefore, it is crucial for Chemical LLMs to understand molecular structures and properties. We manage to acquire large-scale molecule datasets from PubChem~\citep{nihPubChem}, one of the biggest open accessible chemical databases with more than 100M compounds. We include \moleculenum{} molecules along with their notations, descriptions (if applicable), and properties. Besides PubChem, we also leverage the PubChemQC~\citep{nakata2017pubchemqc} dataset, which contains the quantum chemical calculation results of 86M molecules from PubChem, to supplement the quantum chemical properties of these molecules, such as dipole moment and orbital energy. To diversify the final data entry, we randomize the order of the properties and use three different formats: markdown list, markdown table, and JSON dictionary to formulate the molecule data.

\paragraph{Reactions.} Reactions are the major process in the chemical world.
In \ChemFG, we use the reactions from USPTO-FULL~\citep{dai2019gln}, one of the most comprehensive open-sourced chemical reaction databases. To avoid data leakage, we exclude the test set of USPTO-FULL, USPTO-MIT~\citep{jin2017predicting}, and USPTO-50K~\citep{schneider2016s} according to the products of reactions. Moreover, to further enhance the data diversity, we leverage the SMILES~(Simplified Molecular Input Line Entry System) augmentation method introduced in R-SMILES~\citep{zhong2022rsmiles} and achieve a total of 10 times augmentation of data. Finally, a corpus of \reactionnum{} reactions is obtained.

\subsection{Functional Groups Coverage}\label{fgcover}

\begin{figure}
    \centering

    \begin{overpic}[width=\linewidth]{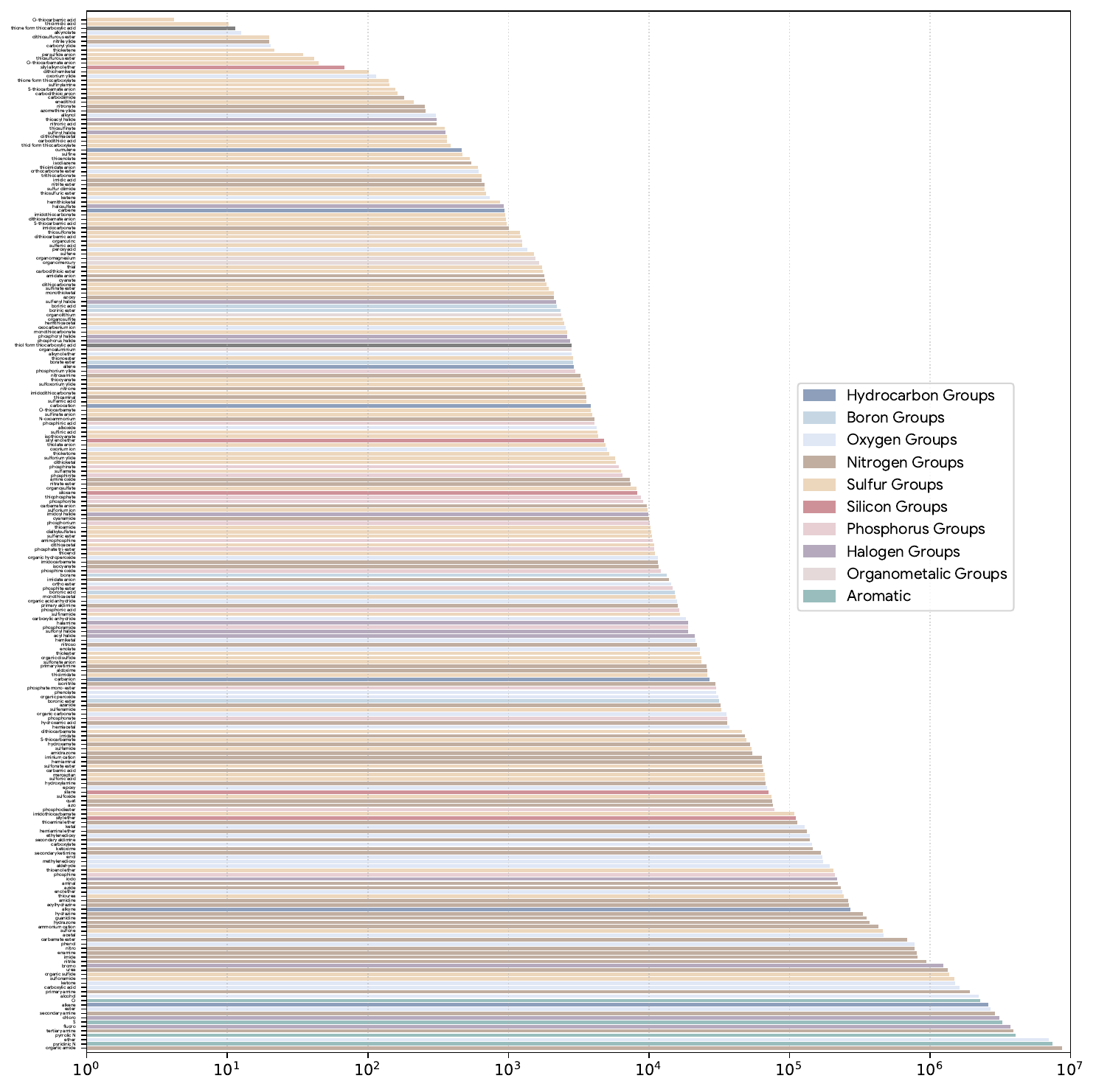}
        \put(55,67){\includegraphics[width=0.4\linewidth]{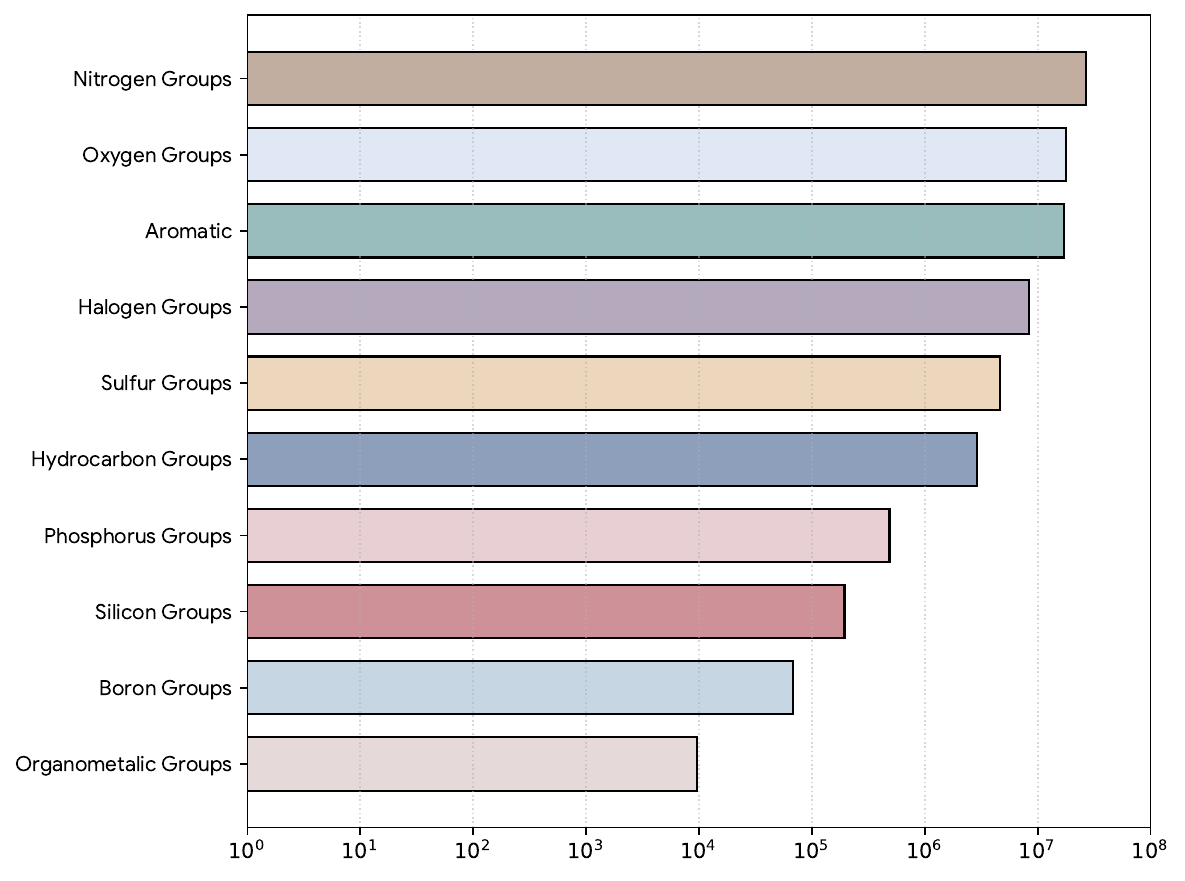}}
    \end{overpic}
    \caption{The distribution of the functional groups in the domain-pretraining corpus.}
    \label{fig:functional}
\end{figure}

The functional groups that can be recognized by our toolkit, ChemFG-Tool, are categorized based on the heteroatoms and listed as follows:
\begin{itemize}
    \item \textbf{Hydrocarbon Groups (7)}: alkene, alkyne, allene, cumulene, carbocation, carbanion, carbene.
    \item \textbf{Boron Groups (6)}: borane, boronic acid, boronic ester, borinic acid, borinic ester, borate ester.
    \item \textbf{Oxygen Groups (36)}: alcohol, alkoxide, ether, phenol, phenolate, enol, enolate, enol ether, alkynol, alkynolate, alkynol ether, ketone, ketene, aldehyde, hemiketal, hemiacetal, ketal, acetal, carboxylic acid, carboxylate, ester, organic acid anhydride, carboxylic anhydride, organic carbonate, organic hydroperoxide, organic peroxide, peroxyacid, ortho ester, orthocarbonate ester, methylenedioxy, ethylenedioxy, oxonium ion, oxocarbenium ion, carbonyl ylide, oxonium ylide, epoxy.
    \item \textbf{Nitrogen Groups (62)}: primary amine, secondary amine, tertiary amine, ammonium cation, quat, amine oxide, enamine, hydroxylamine, hemiaminal, hemiaminal ether, thioaminal, thioaminal ether, aminal, primary ketimine, secondary ketimine, primary aldimine, secondary aldimine, amidine, guanidine, ketoxime, aldoxime, hydrazone, organic amide, amidate anion, imide, carbamic acid, carbamate ester, carbamate anion, azide, azo, hydrazine, acylhydrazine, amidrazone, cyanate, isocyanate, nitrile, isonitrile, cyanamide, carbodiimide, nitrate ester, nitrite ester, nitro, nitroso, nitrosamine, iminium cation, nitrone, nitronic acid, imidic acid, imidate anion, imidate, imidocarbonate, imidocarbamate, urea, azoxy, N-oxoammonium, hydroxamic acid, hydroxamate, azanide, azomethine ylide, nitrile ylide, isodiazene, nitronate.
    \item \textbf{Sulfur Groups (85)}: mercaptan, thiolate anion, organic sulfide, thioenol, enedithiol, thioenolate, thioenol ether, persulfide anion, organic disulfide, sulfenic acid, sulfenic ester, sulfenamide, sulfoxide, sulfone, sulfine, sulfene, sulfinylamine, sulfur diimide, sulfinic acid, sulfonic acid, sulfinate ester, sulfonate ester, sulfinate anion, sulfonate anion, thiosulfinate, thiosulfonate, thiosulfurous ester, dithiosulfurous ester, thiosulfuric ester, organosulfite, organosulfate, dialkylsulfates, sulfinamide, sulfonamide, sulfamic acid, sulfamate, sulfamide, thiocyanate, isothiocyanate, thioketone, thioketene, thial, thioamide, thiourea, hemithioketal, hemithioacetal, dithiohemiketal, dithiohemiacetal, monothioketal, monothioacetal, dithioketal, dithioacetal, carbothioic S-acid, carbothioic O-acid, thiol form thiocarboxylate, thione form thiocarboxylate, thiolester, thionoester, carbodithioic acid, carbodithioic anion, carbodithioic ester, monothiocarbonate, xanthic acid, xanthate, xanthate anion, dithiocarbonate, trithiocarbonate, O-thiocarbamic acid, S-thiocarbamic acid, O-thiocarbamate, S-thiocarbamate, O-thiocarbamate anion, S-thiocarbamate anion, thioimidic acid, thioimidate anion, thioimidate, dithiocarbamic acid, dithiocarbamate, dithiocarbamate anion, imidothiocarbonate, imidodithiocarbonate, imidothiocarbamate, sulfonium ion, sulfonium ylide, sulfoxonium ylide.
    \item \textbf{Silicon Groups (5)}: silane, siloxane, silyl ether, silyl enol ether, silyl alkynol ether.
    \item \textbf{Phosphorus Groups (17)}: phosphine, phosphonium, aminophosphine, phosphine oxide, phosphinic acid, phosphinate, phosphonic acid, phosphonate, phosphite ester, phosphinite, phosphonite, phosphodiester, phosphate mono-ester, phosphate tri-ester, phosphoramide, thiophosphate, phosphonium ylide.
    \item \textbf{Halogen Groups (14)}: fluoro, chloro, bromo, iodo, halamine, sulfenyl halide, sulfinyl halide, sulfonyl halide, halosulfate, phosphoryl halide, phosphorus halide, acyl halide, imidoyl halide, thioacyl halide.
    \item \textbf{Organometalic Groups (5)}: organolithium, organomagnesium, organoaluminium, organozinc, organomercury.
    \item \textbf{Aromatic (4)}: pyrrolic N, pyridinic N, aromatic O, aromatic S.
\end{itemize}

The occurrence of these functional groups in the domain-pretraining corpus is shown in Figure~\ref{fig:functional}.

\subsection{Quality Control}\label{app:anno}

\begin{figure}
    \centering
    \includegraphics[width=0.9\linewidth]{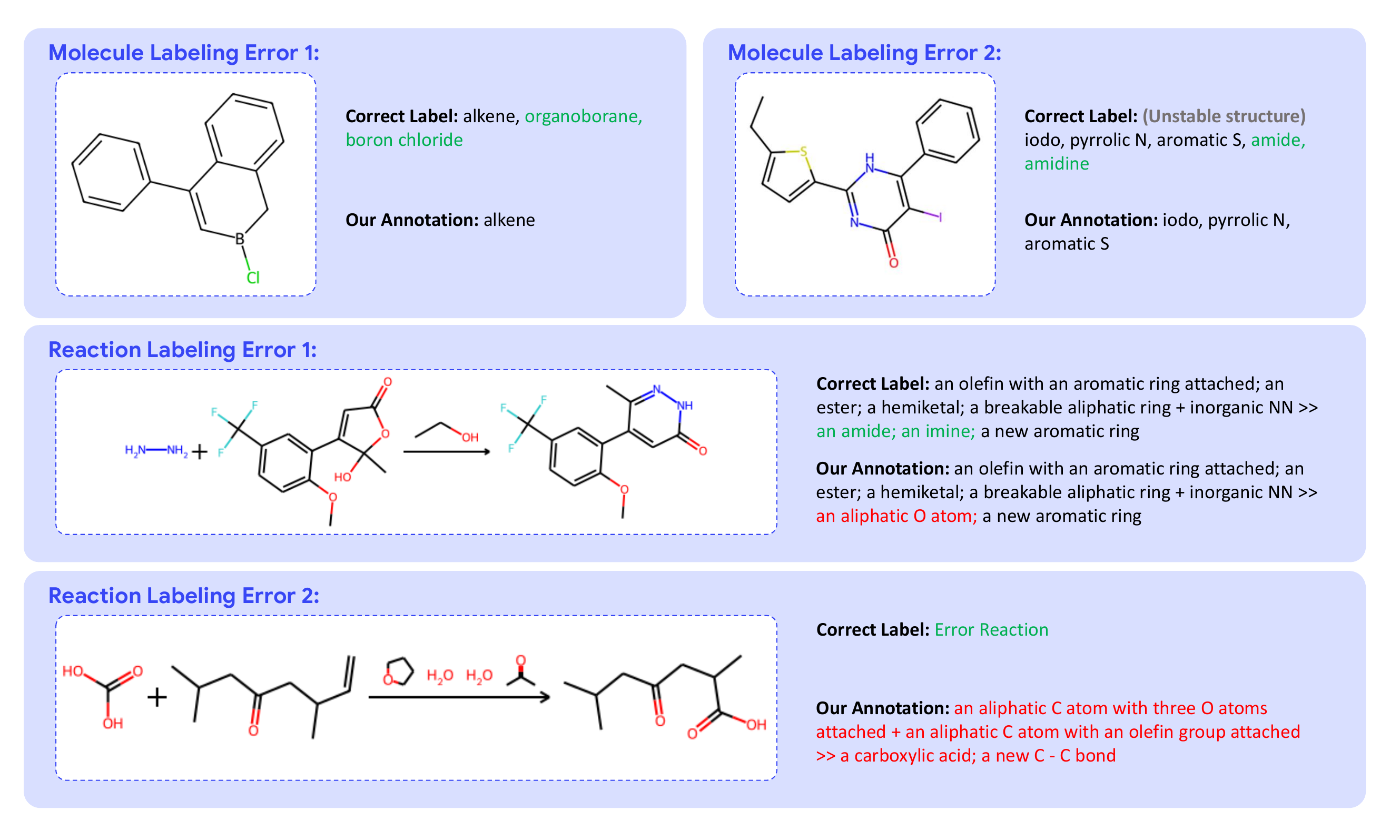}
    \caption{Examples of the error annotations of our functional group identification toolkit, ChemFG-Tool.}
    \label{fig:errors}
\end{figure}

To validate the correctness of our functional group identification toolkit, ChemFG-Tool, we hired three graduate-level chemical experts to conduct manual inspections. Specifically, we sampled 100 annotated molecules and reactions, respectively, and asked the experts to determine whether the annotations were correct. Results show that ChemFG-Tool achieves 98\% accuracy rate on molecules and 89\% on reactions. Examples of the errors are demonstrated in Figure~\ref{fig:errors}.

\section{Instruction Tuning Dataset}\label{sftdata}

\subsection{Raw Data Collection}

\begin{figure*}[!h]
    \centering
    \includegraphics[width=0.6\linewidth,trim=80 0 80 0,clip]{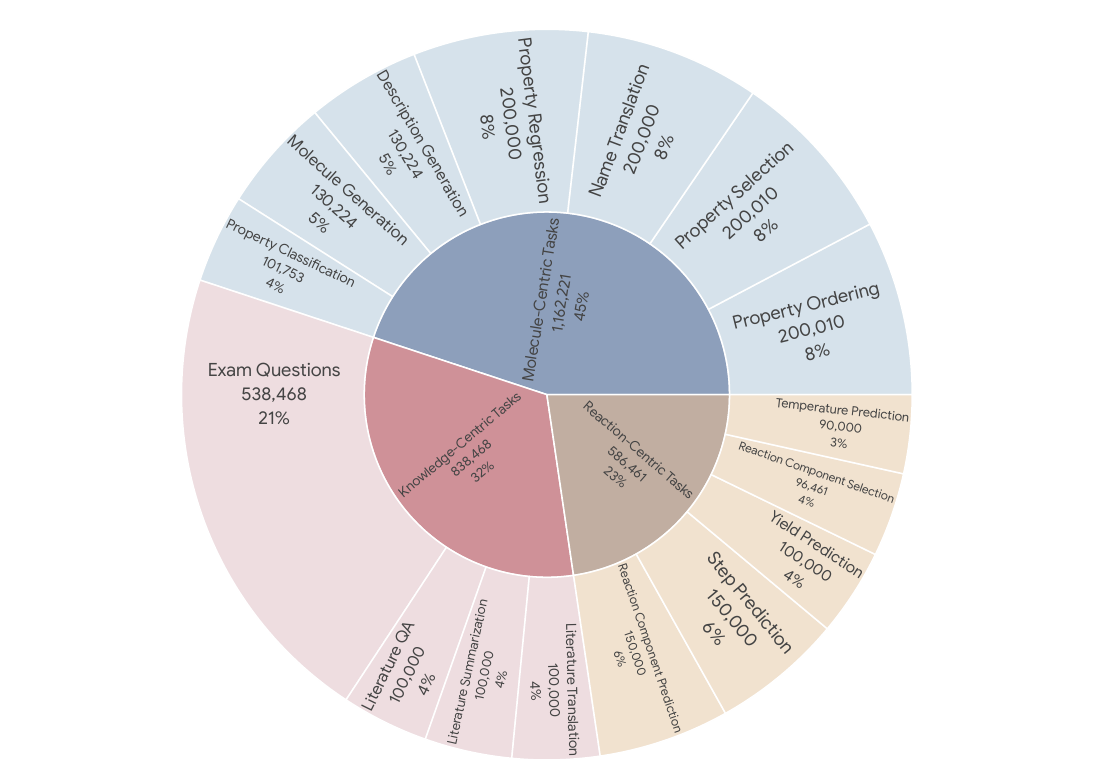} 
    \caption{The distribution of instruction tuning data.}
    \label{fig:dist_sft}
\end{figure*}

Our instruction tuning dataset is constructed of three parts corresponding to the three main information carriers in chemistry: molecule-centric tasks, reaction-centric tasks, and knowledge-centric tasks. The distribution of instruction tuning data is shown in Figure~\ref{fig:dist_sft}.

\subsubsection{Molecule-Centric Tasks}
\begin{itemize}
    \item \textbf{Name Translation}: The name translation between SMILES, IUPAC name, and molecular formula. The data is constructed from PubChem~\citep{nihPubChem}.
    \item \textbf{Description Generation}: The molecule description task is to describe the molecule given its SMILES. The data is constructed from PubChem. We only use the high-quality descriptions that contain more than two sentences.
    \item \textbf{Molecule Design}: The molecule design task is the reverse task of molecule description. It requires the model to predict the SMILES given the molecule description. We use the same high-quality description data from PubChem to construct this task.
    \item \textbf{Property Classification}: These tasks require models to predict the value of molecular properties from a list of candidates (usually yes and no). The data is constructed from 5 of the most popular property classification datasets in MoleculeNet~\citep{wu2018moleculenet}, namely BACE, BBBP, ClinTox, HIV, and Tox21.
    \item \textbf{Property Regression}: These tasks require the models to predict the value of molecular properties, which is a real number. Data are also from MoleculeNet, namely FreeSolv, Lipo, and QM9.
    \item \textbf{Property Ordering}: Provided a list of molecules, models are asked to rank them in ascending or descending order of some specific property. Raw data comes from the same source as property regression.
    \item \textbf{Property Selection}: Provided a list of molecules, models are asked to select the one with the highest or lowest value of some specific property. Raw data comes from the same source as property regression.
\end{itemize}
\subsubsection{Reaction-Centric Tasks}
\begin{itemize}
    \item \textbf{Reaction Completion}: Given an incomplete reaction, models need to complete the missing reactants, reagents, or products. Raw data comes from USPTO-Full~\citep{dai2019gln}, USPTO-MIT~\citep{jin2017predicting}, and USPTO-50K~\citep{schneider2016s}.
    \item \textbf{Step Prediction}: Given a reaction, models are required to predict the experimental procedure to conduct it in the laboratory. Raw data comes from USPTO~\citep{dai2019gln}.
    \item \textbf{Yield Prediction}: In this task, models are required to predict the yield of the given reactions. The data is constructed from the USPTO dataset.
    \item \textbf{Temperature Prediction}: In this task, models are required to predict the temperature that is suitable for the given reactions to conduct. The data is constructed from the USPTO dataset.
    \item \textbf{Reaction Component Selection}: In this task, a series of reactants and reagents is given with a list of candidate molecules. Models need to pick from the candidates the molecules that could participate in the reaction and lead to the highest yield. The data is constructed from the USPTO dataset.
\end{itemize}
\subsubsection{Knowledge-Centric Tasks}
\begin{itemize}
    \item \textbf{Exam Questions}: This task is composed of questions from the exams in middle school and high school. Raw data comes from the Open Internet.
    \item \textbf{Literature QA}: In this task, models are required to answer questions based on the given paragraph. The data is extracted from the long paragraph following the method in SciQAG~\citep{wan2024sciqagframeworkautogeneratedscience}. The raw data comes from the articles in the domain-pertaining. The articles are split into sections and then truncated into paragraphs within 2k to 3k tokens. We ask GPT-4o-mini to extract 15 keywords from each paragraph, then generate 10 question-answer pairs according to them. We adopt another LLM, Qwen2.5-14B-Instruct, to evaluate the quality of the QA pair in 4 dimensions: completeness, accuracy, reasonableness, and agnosticism. The LLM will score the QA pair from 1 to 5 using the designed prompts. QA pairs with any scores below 5 are discarded. If there are more than 1 QA pair left, the questions are asked in conversation turns.
    \item \textbf{Literature Summarization}: In this task, models are required to give a summarization of the paragraph. The summarization is generated from GPT-4o-mini from the paragraph sample.
    \item \textbf{Literature Translation}: In this task, models are required to translate the English paragraph into Chinese. The translation is generated from GPT-4o-mini from the paragraph sample. Since the source data consists of OCR text extracted from English articles, which is inherently noisy, we decided to discard the reverse task of translating Chinese paragraphs into English.
\end{itemize}

\subsection{Instruction Generation}

To acquire a higher generalization capability, we adopt a two-stage process to obtain as diverse a set of instructions as possible for each task. Specifically, based on the number of data entries for each task, we first manually write 5-20 seed task descriptions accordingly. Then, we ask three different models, Qwen2.5-72B-Instruct, Llama-3.1-70B-Instruct, and GPT-4o-mini, to diversify these task descriptions. Specifically, during each request, we sample 5 descriptions from all the generated descriptions and ask the model to generate 10 new descriptions using 5 different prompts one by one. Following this, we append to each of the descriptions the instructions that introduce the input of each data entry by a formatted string of Python and ask GPT-4o-mini to further diversify them. As a result of the above diversification process, we constructed a number of instructions equivalent to approximately 2\% of its total data volume for each task. In other words, each instruction appears around 50 times in the instruction-tuning dataset.

\section{Quality Control of the Distillation Rationale Generation}

\subsection{Comparison of Rationales Generated by Teacher Models}\label{distillation}

\begin{figure}
    \centering
    \includegraphics[width=\linewidth]{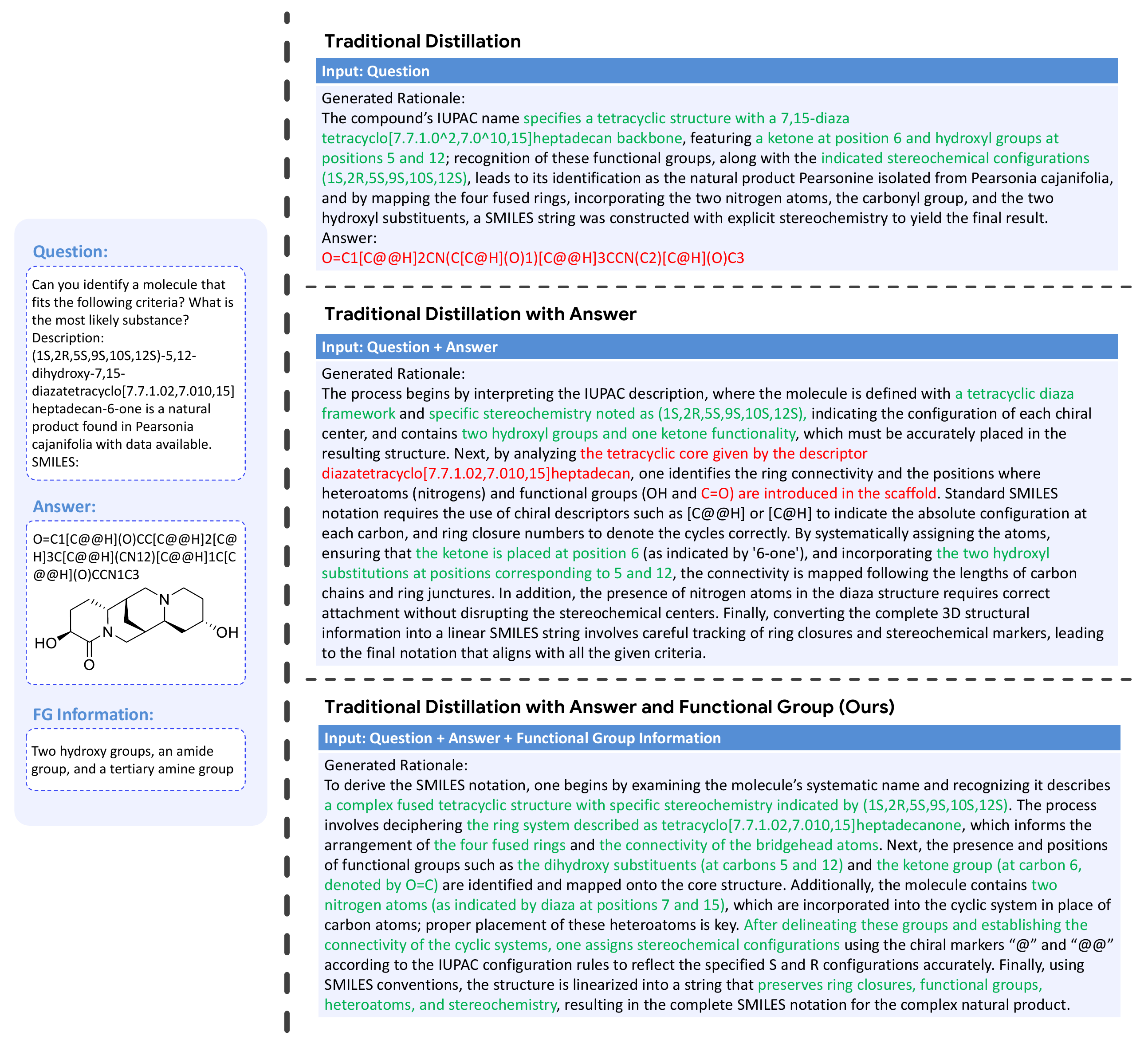}
    \caption{Comparison of rationales generated by o3-mini with and without additional input information. We mark the correct analyses in the rationale as \textcolor{green}{green}, while the wrong ones as \textcolor{red}{red}.}
    \label{fig:dist_ration_comp_a1}
\end{figure}

\begin{figure}
    \centering
    \includegraphics[width=\linewidth]{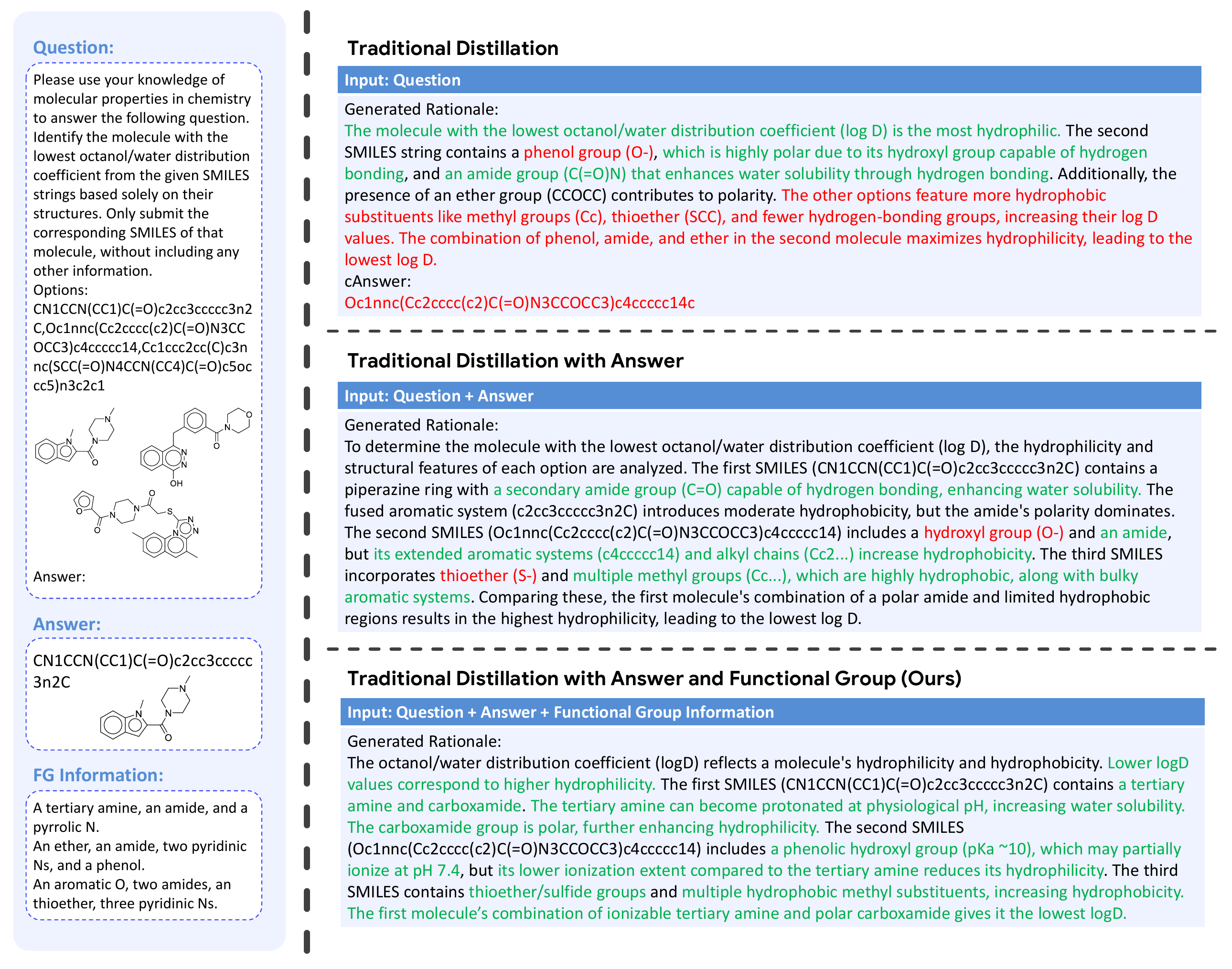}
    \caption{Comparison of rationales generated by DeepSeek-R1 with and without additional input information. We mark the correct analyses in the rationale as \textcolor{green}{green}, while the wrong ones as \textcolor{red}{red}.}
    \label{fig:dist_ration_comp_a2}
\end{figure}

As shown in Figure~7,~\ref{fig:dist_ration_comp_a1}, and \ref{fig:dist_ration_comp_a2}, the rationales generated using our method exhibit distinct advantages in terms of accuracy, completeness, and logicality.

\paragraph{The advantage in accuracy is mainly reflected in Figure~7.} In Figure~7, o3-mini completely misunderstands the chemical reaction that occurs between the given molecules. This might result from its incorrect identification of the reagents and the structure of the complex reactants (these errors are also reflected in the rationales). When given the correct answer, o3-mini still mistakenly identified chloro-ketone as acyl chloride and consistently adhered to this error throughout the reasoning process, resulting in a series of related structural inaccuracies. On the contrary, when the functional group information is provided, o3-mini manages to generate a near-perfect rationale with zero error.

\paragraph{The advantage in completeness is mainly reflected in Figure~\ref{fig:dist_ration_comp_a1}.} In Figure~\ref{fig:dist_ration_comp_a1}, given only the question, DeepSeek-R1 only generates a brief analysis on the second option while dismissing the other three options with a single sentence in total. This overly simple analysis leads to a wrong prediction. With the help of the ground truth answer, the generated rationale analyzes all options individually. However, due to its lack of chemical knowledge, the analysis still exhibits errors in functional group recognition or overlooks key influencing factors. After enhancing chemical knowledge with the functional group information, DeepSeek-R1 finally manages to generate a more comprehensive analysis with few errors.

\paragraph{The advantage in logicality is mainly reflected in Figure~\ref{fig:dist_ration_comp_a2}.} In Figure~\ref{fig:dist_ration_comp_a2}, with only the question, o3-mini can hardly generate any useful rational. The rational merely repeats the IUPAC components mentioned in the question before rushing to a highly inaccurate conclusion without substantive analysis. When given the ground truth answer, o3-mini can construct a reasonably good rationale with minimal factual error. However, the rationale still contains non-negligible issues in terms of logical coherence. A sound reasoning process should follow the approach exemplified by the reasoning chain generated by o3-mini using our method: analyzing in the order of molecular skeleton, functional groups, heteroatoms, and chiral centers. This sequence reflects a step-by-step refinement from the fundamental molecular structure to more intricate structural details. However, with only the question and answer, the generated rationale mixes these analytical steps and lacks critical details, such as the precise position of the nitrogen atom, resulting in a disorganized and incomplete reasoning process.

\subsection{Quality Control}\label{app:dist}

\begin{figure}
    \centering
    \vspace{-10mm}
    \includegraphics[width=0.5\linewidth,trim=80 50 80 50,clip]{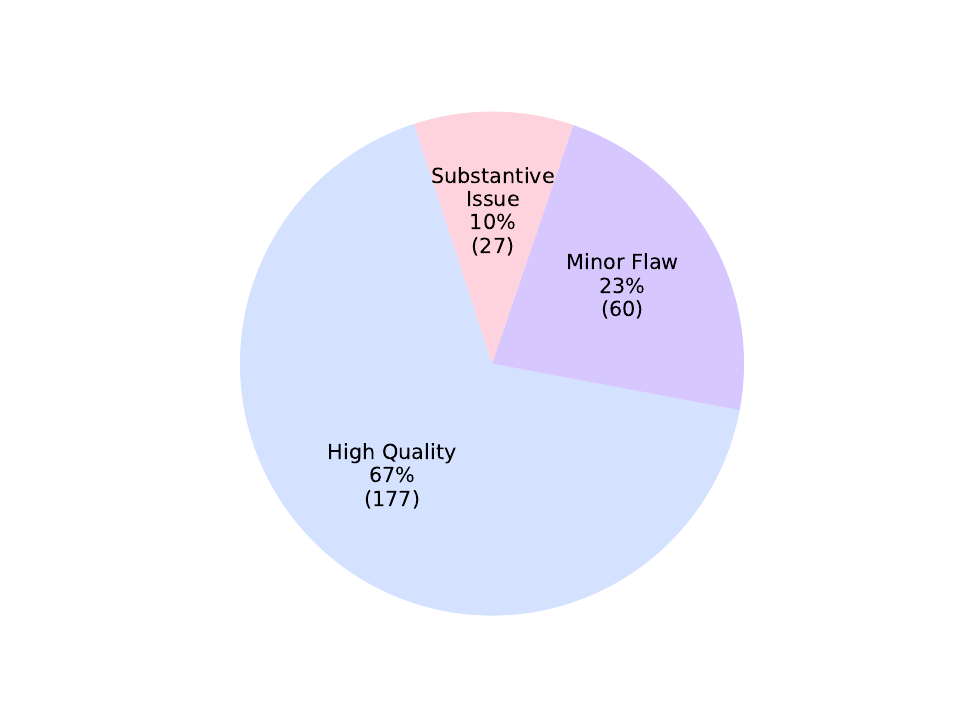}
    \vspace{3mm}
    \caption{Human validation result of teachers' rationales. ``Minor Flaw'' represents acceptable flaws, such as skipping reasoning steps or missing possibilities. ``Substantive Issue'' means severe logic errors or nonsense reasoning.}
    \label{rat_valid}
\end{figure}

\begin{figure}
    \centering
    \includegraphics[width=0.9\linewidth]{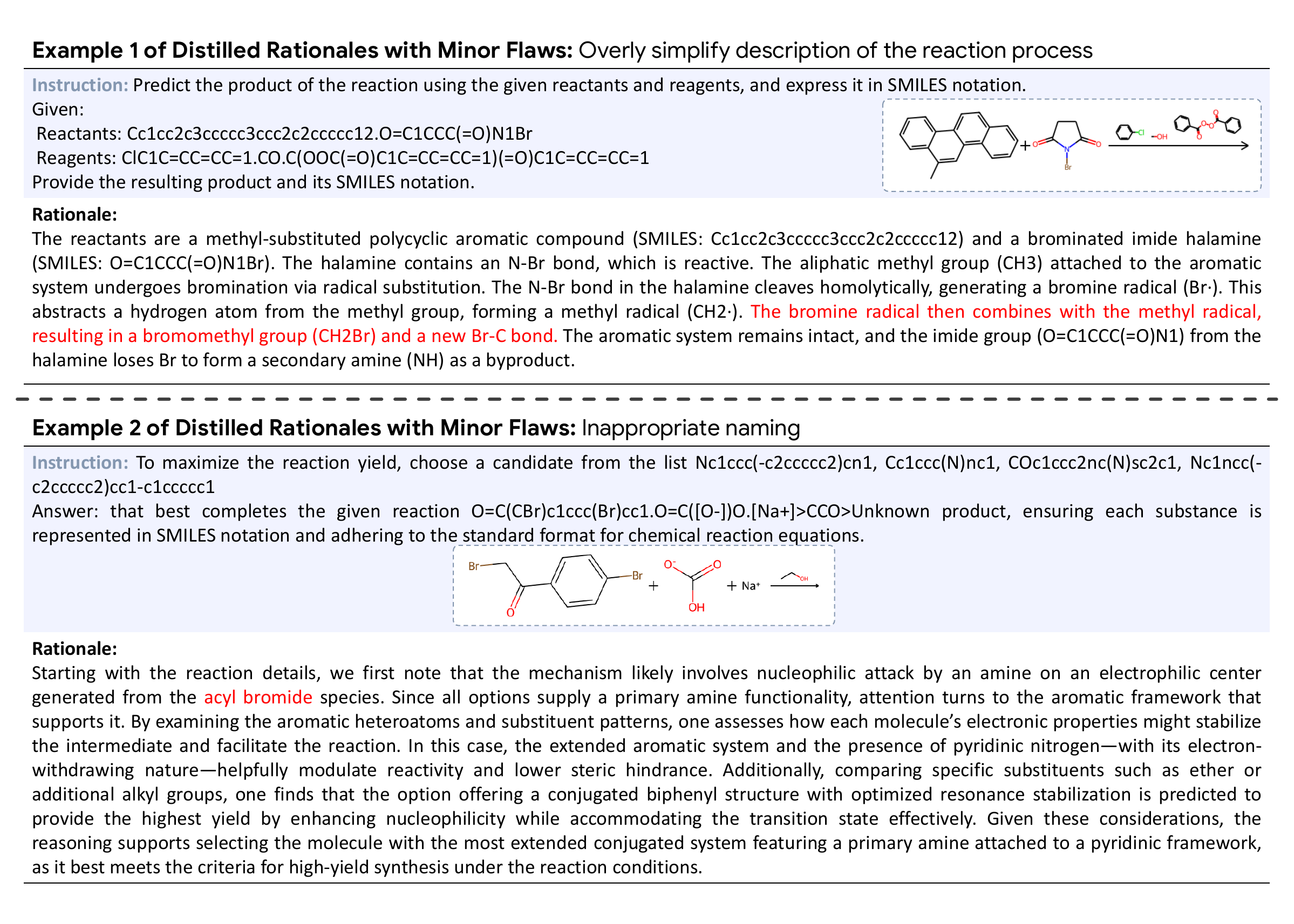}
    \caption{Examples of the distilled rationales with minor flaws.}
    \label{fig:errors1}
\end{figure}

\begin{figure}
    \centering
    \includegraphics[width=0.9\linewidth]{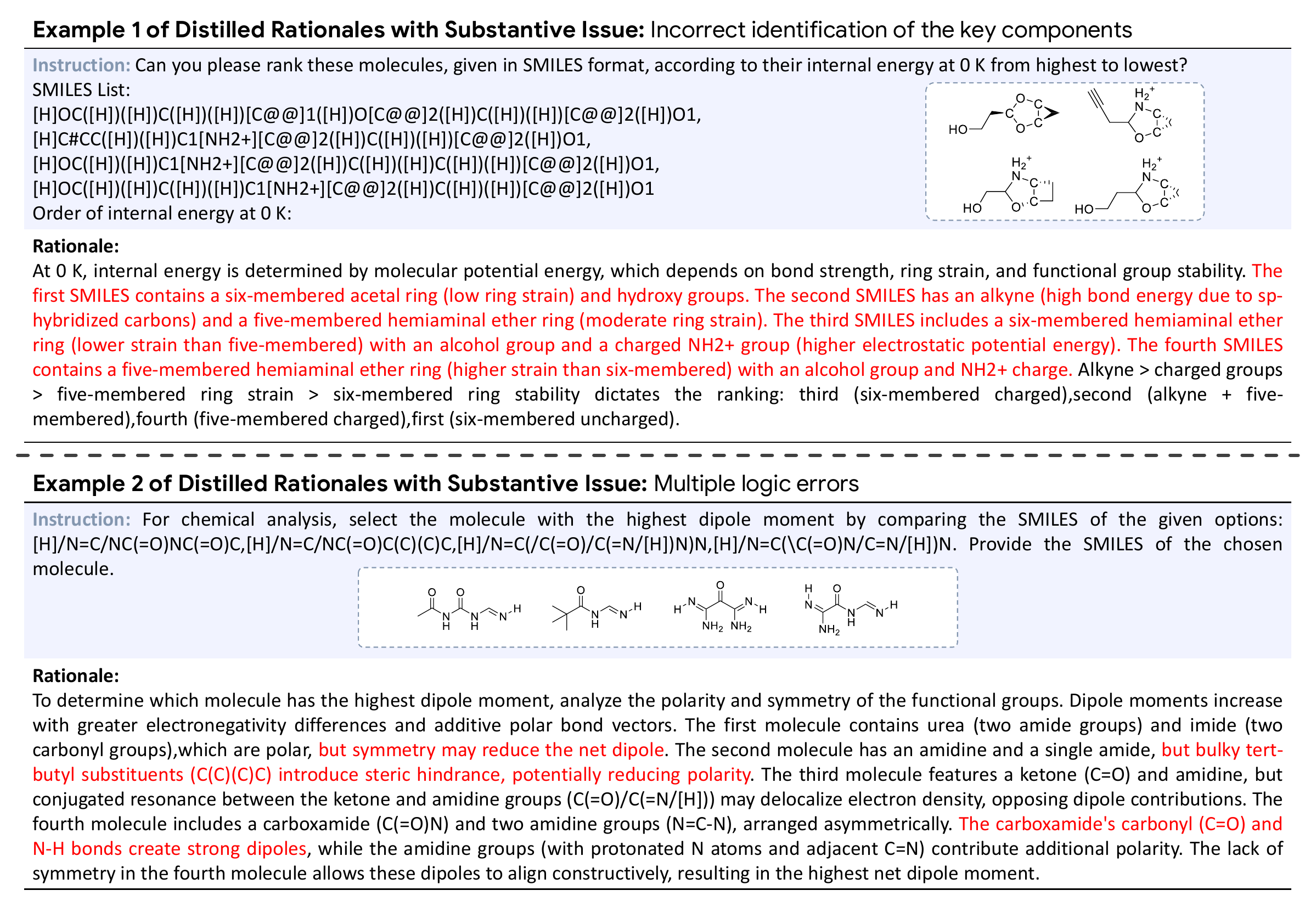}
    \caption{Examples of the distilled rationales with substantive issues.}
    \label{fig:errors2}
\end{figure}

To quantitatively validate the quality of the teachers' rationales generated by our method, we hired three graduate-level chemical experts to perform manual assessments.
Results in Figure~\ref{rat_valid} show that among the sampled 264 rationales, 177 of them (67\%) exhibit sufficiently high quality, 60 of them (23\%) have minor, acceptable flaws such as reasoning step skipping or missing possibilities, and 27 of them (10\%) contain substantive issues such as logic errors or nonsense reasoning. 
Two examples of distilled rationales with minor flaws are shown in Figure~\ref{fig:errors1}, and two examples of distilled rationales with substantive issues are shown in Figure~\ref{fig:errors2}.
Considering that reinforcement learning will be used after distillation to correct errors and improve performance, and that it is difficult to systematically distinguish these problems in the rationales, we used all the data during the distillation process.

\section{Training Details of ChemDFM-R}\label{app:training}

\subsection{Mitigating Data Leakage}

Data leakage is a crucial problem for the evaluation of LLMs. During the training of ChemDFM-R, measures have been taken to reduce the risk of data leakage. For the domain pretraining corpus, we avoid using the same molecules and reactions as those presented in SciKnowEval~\citep{feng2024sciknoweval} and ChemEval~\citep{huang2024chemeval} benchmarks. For instruction tuning and mixed-source distillation, the same molecules and reactions are deliberately excluded from a task if they appear in relative tasks in SciKnowEval and ChemEval. We use canonicalized SMILES to determine whether two molecules are the same. Reactions are considered the same if they share the same product.

\subsection{Statistics about Computation and Data}

We trained ChemDFM-R on NVIDIA A800 Tensor Core GPUs for a total of 30840 GPU hours. Specifically:
\begin{itemize}
    \item[1).] for domain pretraining, the model was trained on the \totaltoken{} ChemFG corpus~(Section \ref{app:chemfg}) for 24728 GPU hours; 
    \item[2).] for instruction tuning, the model was trained on over 7.5 million instructions~(Section \ref{sftdata}), which consist of 2.5 million chemical instructions and 5 million general instructions, for 3785 GPU hours;
    \item[3).] for mixed-sourced distillation, the model was trained on the mixed-sourced distillation dataset, which is of the same scale as the instruction-tuning dataset, for 2059 GPU hours;
    \item[4).] for reinforcement learning, the model was trained on 121,811 samples for 268 GPU hours.
\end{itemize}

\subsection{Hyperparameters and System Prompts}

\paragraph{Training Hyperparameters.} The hyperparameters used during the training of ChemDFM-R are listed in Table~\ref{tab:hyper}.

\paragraph{Inference Setting.} During the inference of ChemDFM-R, we set the temperature to 0.6, topK to 1, and topP to 1 with no other penalties.



\paragraph{System Prompts.} During the training of ChemDFM-R, three different system prompts are used. Specifically:

\begin{itemize}
    \item For all samples in the instruction-tuning dataset and the non-reasoning samples in the mixed-source distillation dataset, we use: "You are a helpful assistant."
    \item For the pseudo-reasoning data in the mixed-source distillation dataset, we use: "You are a helpful assistant that is good at answer chemical questions. You will analyze the presence of functional groups in molecules and the changes in functional groups during reactions before giving response. These analyses will help you solve the problem better. The analyses and answer are enclosed within <think> </think> and <answer> </answer> tags, respectively.$\backslash$ni.e.,$\backslash$n<think>$\backslash$nanalyses here$\backslash$n</think>$\backslash$n<answer>$\backslash$nanswer here$\backslash$n</answer>"
    \item For the distilled data and the training of reinforcement learning, we use "You are a helpful assistant that is good at reasoning. You always reason thoroughly before giving a response. The reasoning process and answer are enclosed within <think> </think> and <answer> </answer> tags, respectively.$\backslash$ni.e.,$\backslash$n<think>$\backslash$nreasoning process here$\backslash$n</think>$\backslash$n<answer>$\backslash$nanswer here$\backslash$n</answer>"
\end{itemize}

\begin{table}[]
    \centering
    \caption{The training hyperparameters used during the training of ChemDFM-R.}
    \begin{tabular}{c|cccc}
        \toprule
        & Domain & Instruction & Mixed-source & Reinforcement \\
        & Pretraining & Tuning  & Distillation & Learning \\
        \midrule
        Initial Learning Rate & 1e-5 & 1e-5 & 1e-5 & 5e-7 \\
        Minimal Learning Rate & 1e-6 & 0 & 0 & 0 \\
        Optimizer & \multicolumn{4}{c}{Adam(0.9, 0.95)} \\
        Scheduler & \multicolumn{4}{c}{Cosine} \\
        Max Sequence Length & 8192 & 8192 & 8192 & - \\
        Max Generation Length & - & - & - & 1024 \\
        Train Batch Size & 624 & 512 & 512 & 128 \\
        Rollout Batch Size & - & - & - & 512 \\
        Epochs & 1 & 2 & 1 & 1 \\
        DAPO Group Size & - & - & - & 8 \\
        DAPO Epsilon & - & - & - & (0.2, 0.3) \\
        Initial KL Coefficient & - & - & - & 1e-3 \\
        
        \bottomrule
    \end{tabular}
    \label{tab:hyper}
\end{table}

\section{The Analysis of ChemDFM-R's Rationale}\label{more_rationale}

\begin{figure}
    \centering
    \includegraphics[width=0.8\linewidth]{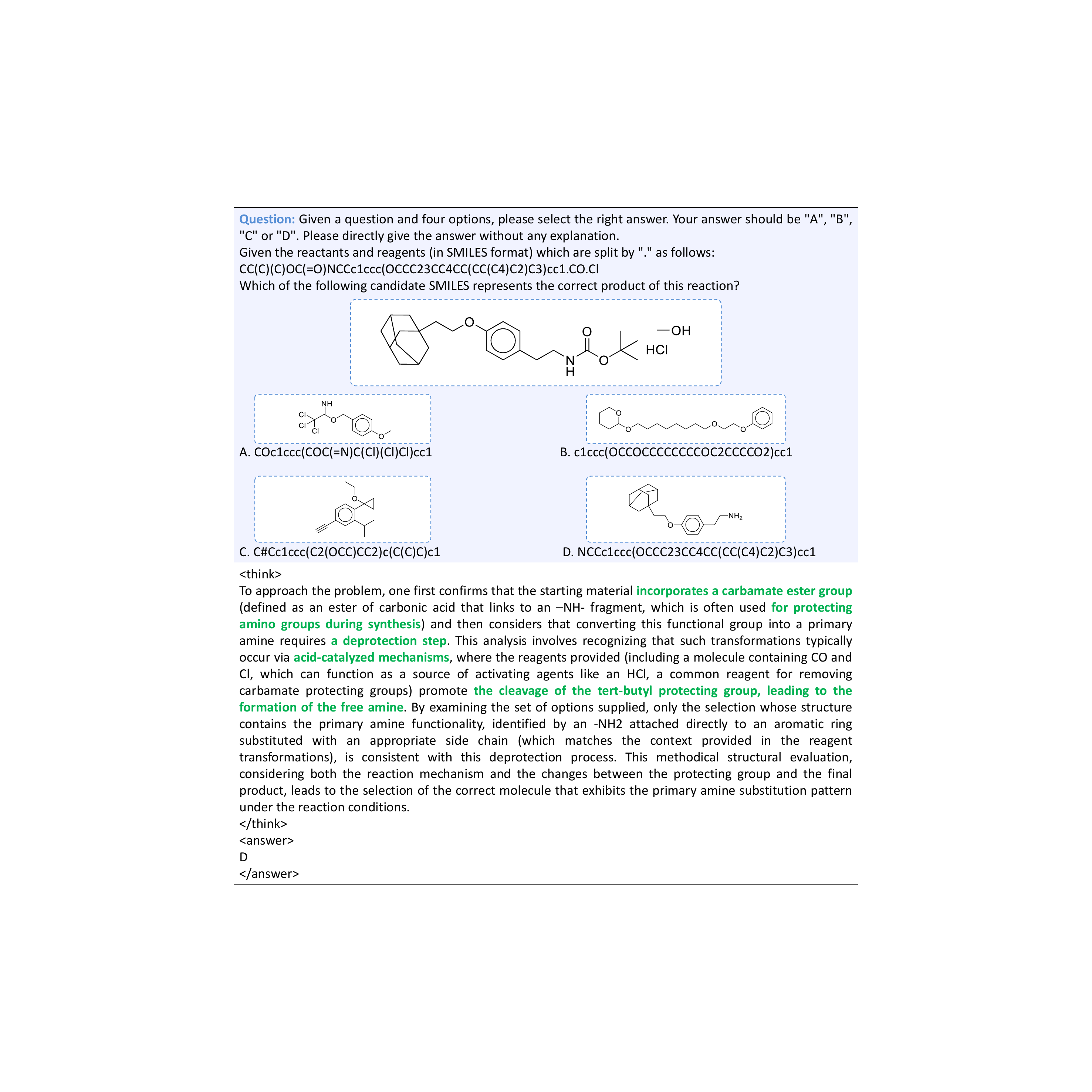} 
    \caption{An example of the responses generated by ChemDFM-R on the reaction prediction task.}
    \label{fig:rationale}
\end{figure}
\begin{figure}
    \centering
    \includegraphics[width=0.8\linewidth]{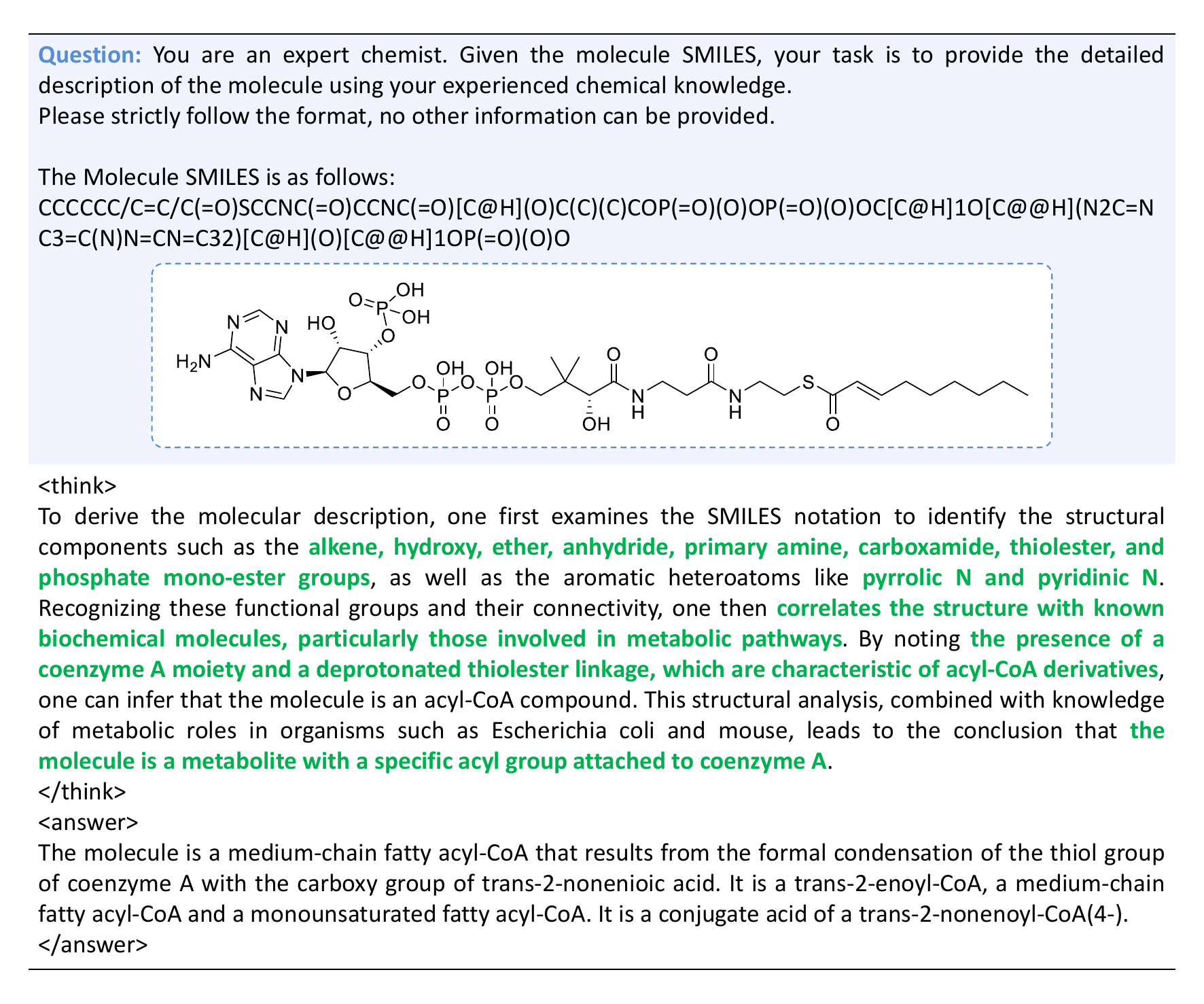} 
    \caption{An example of the responses generated by ChemDFM-R on the molecule captioning task.}
    \label{fig:rationale_1}
\end{figure}
\begin{figure}
    \centering
    \includegraphics[width=0.8\linewidth]{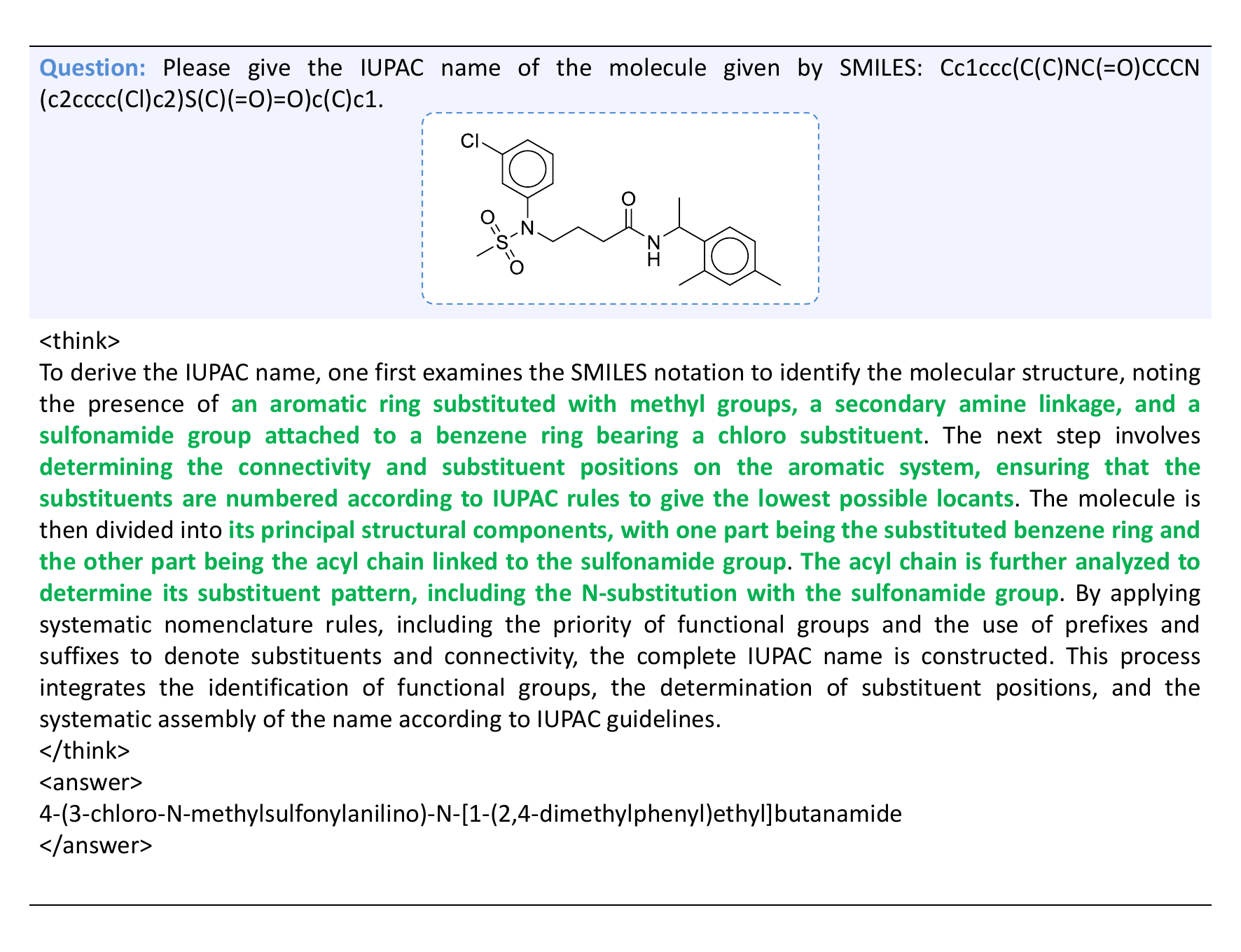} 
    \caption{An example of the responses generated by ChemDFM-R on the SMILES to IUPAC task.}
    \label{fig:rationale_2}
\end{figure}

In Figure~\ref{fig:rationale}, ChemDFM-R is asked to pick a correct product for the given reaction. Instead of wasting time (tokens) on analyzing the SMILES in great detail, which DeepSeek-R1 always does, ChemDFM-R1 goes directly into the key point of this question: the functional groups present in the reactants and the potential reactions between them. Specifically, ChemDFM-R successfully identifies the key functional group, the carbamate ester group. By recalling that the carbamate ester group is typically used to protect amino groups, ChemDFM-R infers that the reaction taking place is likely a deprotection reaction. Then, ChemDFM-R confirms its assumption by examining the provided reagents. Finally, ChemDFM-R predicts the feature of the possible product and picks the option that matches it. This example demonstrates the precision of ChemDFM-R in finding the key point of chemical questions, and the effectiveness and efficiency of ChemDFM-R's rationales. Moreover, it is also worth noticing that, instead of using the ``elimination-shortcut'' which is commonly adopted by other cutting-edge reasoning LLMs, ChemDFM-R directly reasoned out the reaction mechanism and the features of the correct answer, thereby selecting the correct option.

As illustrated in Figure~\ref{fig:rationale_1}, when asked to describe a molecule given by SMILES, ChemDFM-R first analyzes the functional groups present in the molecule, such as the alkene group, the phosphate mono-ester group, and the pyrrolic N group. Then, ChemDFM-R successfully correlates the composition and connectivity of these functional groups with metabolic pathways and further manages to identify the molecule as a coenzyme A derivative. After that, it recognizes the deprotonated thiolester linkage in the molecule and further narrows down the molecule to an acyl-CoA derivative. Finally, ChemDFM-R gives a relatively comprehensive description of the molecule. ChemDFM-R even provides the potential role of this molecule in metabolic processes in its rationale, further demonstrating its strong reasoning ability as well as the value of the rationale as a complement to the final answer.

Figure~\ref{fig:rationale_2} showcases an example response of ChemDFM-R when asked to generate the IUPAC name of the given molecule. The IUPAC name is the standard name for a molecule, assigned according to the rules established by the International Union of Pure and Applied Chemistry (IUPAC). It can effectively reflect the functional groups present in the molecule and their connectivity. Therefore, ChemDFM-R starts its reasoning with a comprehensive analysis of the functional groups of the given molecule. Then, it emphasizes the importance of correctly labeling the atoms, which is precisely an area where large language models are particularly prone to errors. After this, ChemDFM-R follows the rule of IUPAC naming and divides the molecule into its principal structural components. It also specifically points out the N-substitution with the sulfonamide group. Finally, a complicated and correct IUPAC name is predicted by ChemDFM-R.

\section{Detail Results of Benchmark Evaluation}\label{scores}

\subsection{ChemEval}

We consider the L1 and L2 level tasks in ChemEval~\citep{huang2024chemeval} as text-centric tasks, the L3 level tasks as molecule-centric tasks, and the L4 level tasks as reaction-centric tasks. Moreover, there are tasks that we can not achieve a feasible grading in ChemEval. We temporarily skip these tasks. The raw results are demonstrated in Table~\ref{tab:chem}.

\begin{table}[]
    \centering
    \begin{adjustbox}{angle=90,width=\linewidth}
    \begin{tabular}{c|c|ccccc|cc|ccc|c}
    \toprule
    \multirow{2}{*}{Tasks} & \multirow{2}{*}{Metric} & \multirow{2}{*}{MolInst} & ChemLLM & ChemDFM & ChemDFM & \multirow{2}{*}{ether0} & \multirow{2}{*}{GPT-4o} & Qwen3-14B & DeekSeek & Qwen3-14B & \multirow{2}{*}{o4-mini} & \multirow{2}{*}{ChemDFM-R} \\
    &&& -20B-DPO & -13B-v1.0 & -8B-v1.5 &&& (no think) & -R1 & (think) && \\
    \midrule
    \rowcolor{grey}\multicolumn{13}{c}{\textit{Text-Centric Tasks}} \\
    MCTask & Acc & 40.0 & 35.0 & 50.0 & 0.0 & 10.0 & 75.0 & 75.0 & \textbf{90.0} & 80.0 & 65.0 & 64.0 \\
    FBTask & Acc & 0.0 & 0.0 & \textbf{5.0} & 0.0 & 0.0 & 0.0 & 0.0 & 0.0 & 0.0 & 0.0 & 1.0 \\
    TFTask & Acc & 55.0 & 75.0 & 70.0 & 5.0 & 0.0 & 80.0 & 85.0 & \textbf{95.0} & 80.0 & 85.0 & 81.0 \\
    SATask & BLEU & 0.6 & \textbf{17.3} & 12.2 & 10.6 & 0.4 & 15.0 & 9.3 & 4.3 & 9.2 & 10.3 & 11.2 \\
    CNER & F1 & 0.0 & 44.1 & 59.0 & 22.8 & 19.2 & 66.7 & 60.9 & \textbf{69.6} & 67.1 & 57.1 & 63.1 \\
    CERC & F1 & 4.5 & 0.0 & 12.3 & 0.0 & 0.0 & 20.3 & 24.1 & \textbf{37.7} & 20.2 & 25.5 & 24.9 \\
    AddE & F1 & 34.2 & 58.7 & 46.7 & 5.0 & 0.0 & 70.0 & 65.0 & 70.8 & 65.0 & 68.3 & \textbf{73.3} \\
    SolvE & F1 & 58.3 & 70.0 & \textbf{82.5} & 25.0 & 5.0 & 75.0 & 73.3 & 79.0 & 54.0 & 72.3 & 76.3 \\
    TempE & F1 & 70.0 & 75.0 & 20.2 & 25.0 & 10.0 & 75.0 & \textbf{95.0} & \textbf{95.0} & 90.0 & \textbf{95.0} & \textbf{95.0} \\
    TimeE & F1 & \textbf{95.0} & \textbf{95.0} & 40.0 & \textbf{95.0} & 5.0 & \textbf{95.0} & \textbf{95.0} & \textbf{95.0} & \textbf{95.0} & \textbf{95.0} & \textbf{95.0} \\
    CharME & F1 & 20.7 & 67.0 & 26.2 & 49.1 & 16.8 & \textbf{79.0} & 73.8 & 71.0 & 60.7 & 76.0 & 61.6 \\
    CatTE & F1 & 80.0 & 90.0 & 55.0 & 30.0 & 0.0 & 95.0 & 85.0 & \textbf{100.0} & 85.0 & 90.0 & 67.0 \\
    YieldE & F1 & 0.0 & 55.0 & 20.0 & 30.0 & 20.0 & 80.0 & 75.0 & 75.0 & \textbf{95.0} & 75.0 & 57.0 \\
    AbsGen & BLEU & 56.3 & 45.7 & 54.0 & 4.4 & 2.1 & 66.8 & \textbf{70.4} & 62.2 & 64.7 & 67.4 & 65.1 \\
    OLGen & BLEU & 18.9 & 12.5 & 22.4 & 2.7 & 8.7 & 14.3 & \textbf{23.6} & 10.0 & 18.0 & 13.1 & 15.4 \\
    TopC & Acc & 40.0 & 50.0 & 40.0 & 20.0 & 0.0 & 50.0 & 40.0 & \textbf{55.0} & 50.0 & 50.0 & 45.0 \\
    ReactTR & F1 & 20.0 & 13.3 & 15.0 & 5.0 & 5.0 & \textbf{35.0} & 30.0 & 25.0 & 30.0 & 20.0 & 28.5 \\
    \midrule
    \rowcolor{grey}\multicolumn{13}{c}{\textit{Molecule-Centric Tasks}} \\
    MolNG & BLEU & 3.6 & 45.5 & 75.9 & 34.6 & 46.7 & 59.9 & 43.1 & 9.9 & 41.7 & 68.6 & \textbf{83.3} \\
    IUPAC2MF & EM & 0.0 & 0.0 & 25.0 & 0.0 & 0.0 & 40.0 & 0.0 & 5.0 & 0.0 & 60.0 & \textbf{73.0} \\
    SMILES2MF & EM & 0.0 & 0.0 & 45.0 & 0.0 & 0.0 & 10.0 & 5.0 & 0.0 & 0.0 & 55.0 & \textbf{86.0} \\
    IUPAC2SMILES & EM & 0.0 & 0.0 & 10.0 & 0.0 & 30.0 & 0.0 & 0.0 & 5.0 & 0.0 & 15.0 & \textbf{55.0} \\
    SMILES2IUPAC & EM & 0.0 & 0.0 & 0.0 & 5.0 & 0.0 & 0.0 & 0.0 & 0.0 & 0.0 & 0.0 & \textbf{32.0} \\
    S2S & EM & 0.0 & 0.0 & 0.0 & 0.0 & 0.0 & 0.0 & 0.0 & 0.0 & 0.0 & \textbf{5.0} & 0.0 \\
    MolPC & Acc & 49.9 & 52.8 & 61.0 & 9.4 & 0.0 & 64.1 & 54.5 & 57.1 & 58.9 & 65.7 & \textbf{72.5} \\
    Mol2PC & BLEU & 29.8 & 36.4 & 38.9 & 18.5 & 2.3 & 37.2 & 40.8 & 30.4 & 38.5 & 33.7 & \textbf{57.8} \\
    \midrule
    \rowcolor{grey}\multicolumn{13}{c}{\textit{Reaction-Centric Tasks}} \\
    SubRec & Acc & 0.0 & 3.3 & 8.2 & 2.0 & 0.0 & 7.5 & 0.0 & 5.0 & 0.0 & 18.3 & \textbf{23.5} \\
    LRec & Acc & 0.0 & 0.0 & 20.1 & 0.0 & 0.0 & 10.0 & 5.0 & 10.0 & 35.0 & 0.0 & \textbf{42.0} \\
    RRec & Acc & 0.0 & 0.0 & 19.0 & 0.0 & 5.0 & 10.0 & 15.0 & 5.0 & 25.0 & 30.0 & \textbf{34.0} \\
    SolvRec & Acc & 0.0 & 0.0 & 5.0 & 0.0 & 0.0 & \textbf{35.0} & 2.0 & 25.0 & \textbf{35.0} & 30.0 & 31.0 \\
    TempRec & RMSE $\downarrow$ & 76.5 & 27.1 & 71.5 & 22.2 & $\infty$ & 21.7 & 23.4 & \textbf{17.9} & 34.6 & 22.2 & 31.8 \\
    TimeRec & RMSE $\downarrow$ & 20.5 & 18.7 & 17.8 & 9.4 & $\infty$ & \textbf{6.6} & 8.0 & 9.5 & 10.7 & 9.6 & 18.8 \\
    PPre & Acc & 0.0 & 0.0 & 3.3 & 5.0 & \textbf{73.3} & 10.0 & 0.0 & 18.2 & 5.8 & 35.0 & 60.0 \\
    YPred & Acc & 20.0 & 20.0 & 20.0 & 0.0 & 0.0 & \textbf{55.0} & 14.9 & 25.0 & 25.0 & 45.0 & 23.0 \\
    RatePred & Overlap & 3.0 & 13.8 & 4.7 & 9.0 & 0.0 & 18.2 & 14.9 & \textbf{31.0} & 9.9 & 7.2 & 14.1 \\
    IMDer & Acc & 0.0 & 0.0 & \textbf{10.0} & 0.0 & 0.0 & 5.0 & 0.0 & 5.0 & 0.0 & 5.0 & 0.0 \\
    \bottomrule
    \end{tabular}
    \end{adjustbox}
    \caption{The detailed benchmark results of different models on ChemEval~\citep{huang2024chemeval}.}
    \label{tab:chem}
\end{table}

As illustrated in Table~\ref{tab:chem}, ChemDFM-R manages to achieve competitive performance in the text-centric tasks compared with the cutting-edge LLMs, while achieving SOTA performance across all the molecule-centric tasks and a large portion of the reaction-centric tasks. A detailed analysis of the task characteristics reveals that ChemDFM-R tends to perform less effectively on tasks involving numerical prediction, which will be a key focus of our future optimization efforts.

\subsection{SciKnowEval}\label{app:sciknoweval}

We group the tasks in SciKnowEval~\citep{feng2024sciknoweval} based on their input and output. Specifically, the task is classified as a text-centric task if there is no SMILES appear in its input or output, as a reaction-centric task if there are reaction SMILES appear in its input or output, and as a molecule-centric task otherwise. Due to budget limit, we currently skip the tasks that require GPT-4o for grading. The raw results are demonstrated in Table~\ref{tab:sci}.

\begin{table}[]
\setlength{\tabcolsep}{3pt}
    \centering
    \begin{adjustbox}{angle=90}
    \begin{tabular}{c|ccccc|cc|ccc|c}
    \toprule
    \multirow{2}{*}{Tasks} & \multirow{2}{*}{MolInst} & ChemLLM & ChemDFM & ChemDFM & \multirow{2}{*}{ether0} & \multirow{2}{*}{GPT-4o} & Qwen3-14B & DeekSeek & Qwen3-14B & \multirow{2}{*}{o4-mini} & \multirow{2}{*}{ChemDFM-R} \\
    && -20B-DPO & -13B-v1.0 & -8B-v1.5 &&& (no think) & -R1 & (think) && \\
    \midrule
    \rowcolor{grey}\multicolumn{12}{c}{\textit{Text-Centric Tasks}} \\
    Chem LiterQA & 75.4 & 78.7 & 77.6 & 79.3 & 57.3 & 85.8 & 84.2 & 86.3 & 84.0 & \textbf{88.9} & 80.2 \\
    React Mech Infer. & 96.3 & 98.5 & 98.1 & 97.8 & 77.7 & \textbf{100.0} & 98.5 & 96.7 & 98.1 & 97.8 & 98.4 \\
    Comp Iden. and Prop. & 93.6 & 97.4 & 95.2 & 96.8 & 72.8 & 98.6 & 98.2 & 96.4 & 98.6 & \textbf{99.0} & 97.3 \\
    Chem DU & 95.2 & 98.2 & 97.4 & 96.5 & 74.4 & \textbf{99.5} & 98.6 & 98.6 & 98.9 & 99.4 & 97.9 \\
    Chem HV & 82.8 & 88.8 & 85.9 & 88.3 & 0.9 & \textbf{94.9} & 93.6 & 93.2 & 92.1 & 93.4 & 92.1 \\
    Chem RI & 92.8 & 95.9 & 94.6 & 95.5 & 60.0 & 96.9 & 97.1 & 95.3 & 96.3 & 96.3 & \textbf{97.2} \\
    Balancing Eq. & 0.6 & 16.3 & 12.3 & 21.7 & 0.0 & 10.5 & 7.7 & 11.2 & 7.25 & 10.5 & \textbf{27.4} \\
    Chem Cal. & 34.9 & 32.7 & 36.8 & 35.3 & 22.7 & 52.4 & 53.9 & 78.8 & 91.4 & \textbf{92.2} & 52.3 \\
    Mol Tox. Pred & 53.7 & 54.3 & 44.3 & 45.2 & 3.9 & 37.4 & 51.8 & \textbf{62.4} & 51.3 & 55.1 & 43.0 \\
    Chem Safe Test & 68.7 & 74.8 & 59.5 & 65.3 & 17.5 & \textbf{85.2} & 81.2 & 81.7 & 83.1 & 80.4 & 81.5 \\
    \midrule
    \rowcolor{grey}\multicolumn{12}{c}{\textit{Molecule-Centric Tasks}} \\
    Mol Name Conv. & 60.3 & 70.8 & 69.7 & 81.0 & 35.1 & 86.7 & 71.5 & 41.8 & 83.7 & \textbf{95.7} & 89.9 \\
    Mol Prop. Pred & 31.8 & 33.9 & 38.1 & 48.0 & \textbf{48.9} & 44.0 & 37.9 & 41.2 & 41.6 & 48.1 & 44.6 \\
    Mol Cap. & 14.4 & 21.0 & 22.7 & 22.5 & 0.9 & 13.9 & 8.6 & 12.6 & 10.0 & 14.0 & \textbf{50.9} \\
    Mol Weight Cal. & 25.7 & 25.9 & 16.9 & 22.9 & 22.4 & 25.9 & 17.0 & 4.4 & 31.1 & \textbf{62.8} & 35.4 \\
    Mol Prop. Cal. & 38.9 & 33.2 & 31.1 & 29.3 & 27.2 & 34.7 & 37.2 & 39.1 & 38.1 & \textbf{45.4} & 18.7 \\
    Mol Stru. Pred & 33.1 & 30.9 & 35.6 & 25.9 & 31.3 & 46.1 & 30.9 & 7.5 & 36.6 & \textbf{57.4} & 31.7 \\
    Mol Gen. & 64.4 & 52.9 & 84.4 & \textbf{89.4} & 70.2 & 60.9 & 38.6 & 69.3 & 44.5 & 75.4 & 86.6 \\
    \midrule
    \rowcolor{grey}\multicolumn{12}{c}{\textit{Reaction-Centric Tasks}} \\
    Reaction Pred & 39.2 & 92.8 & 91.6 & 95.9 & 55.8 & 48.0 & 90.5 & 82.5 & 93.7 & \textbf{99.3} & 98.4 \\
    Retrosynthesis & 44.3 & 83.4 & 78.5 & 77.8 & 76.2 & 41.2 & 70.0 & 83.4 & 83.2 & \textbf{95.2} & 89.4 \\
    \bottomrule
    \end{tabular}
    \end{adjustbox}
    \caption{The detailed benchmark results of different models on SciKnowEval~\citep{feng2024sciknoweval}.}
    \label{tab:sci}
\end{table}

As illustrated in Table~\ref{tab:sci}, ChemDFM-R achieves competitive performance on SciKnowEval compared to cutting-edge LLMs. It is worth noting that ChemDFM-R’s performance advantage is less pronounced on SciKnowEval than on ChemEval. This is primarily because most tasks in SciKnowEval are formulated as multiple-choice questions, which substantially reduce the burden on the model’s comprehension and generation processes, allowing it to arrive at correct answers through ``shortcuts'' such as option comparison and elimination.

\section{Details of the Human Assessments of Rationale Quality}\label{app:human}



The ten original questions we used are listed as follows.

\paragraph{Organic Chemistry:}
\begin{itemize}
    \item \citep{yao2025convergent} I have used m-CPBA to convert the carbon-carbon double bond within the [H][C@]12CC[C@@]3(CC [C@]1C)C(=O)C1=C[C@@]4(C)CC[C@@](C(C)C)[C@]4([H])C[C@]1([H])[C@]23C into an epoxide, and obtained chiral epoxy products with different ratios (d.r. = 5:1). Please propose possible reasons. 
    \item \citep{zhou2025photocatalyst} Clc1ccccc1 is difficult to react with [O-]C(F)(F)F under normal conditions, but it can be converted into a free radical cation under photocatalytic conditions and can react with [O-]C(F)(F)F in the presence of [Ag+] to obtain the product in high yield. Please provide the structure of the product.
\end{itemize}
\paragraph{Inorganic Chemistry:}
\begin{itemize}
    \item \citep{schwarzmann2025methylbismuth} C[Bi+2]([O]1CCCC1)([O]1CCCC1)([O]1CCCC1)([O]1CCCC1)[O]1CCCC1 is a newly reported strong Lewis acid. Please provide the oxidation state, ligand, and coordination number of the metal ion in C[Bi+2]([O]1CCCC1)([O]1CCCC1)([O]1CCCC1)([O]1CCCC1)[O]1CCCC1, and explain the reason why it has strong Lewis acidity.
    \item \citep{mandai2025stable} Fc1c(F)c([B-](c2c(F)c(F)c(B3Oc4ccccc4O3)c(F)c2F)(c2c(F)c(F)c(B3Oc4ccccc4O3)c(F)c2F)c2c(F)c (F)c(B3Oc4ccccc4O3)c(F)c2F)c(F)c(F)c1B1Oc2ccccc2O1 is a newly reported stable Lewis acidic anion, which breaks the previous understanding that anions are incompatible with Lewis acids. Please analyze its structure, explain why it can act as a Lewis acid, and indicate its binding sites with Lewis bases.
\end{itemize}
\paragraph{Materials Chemistry:}
\begin{itemize}
    \item \citep{li2025achieving} What is oxygen evolution reaction (OER)? Please propose a reasonable mechanism of heterogeneous OER under acidic conditions.
    \item \citep{liu2025harvesting} C1=Cc2cc3ccc(cc4nc(cc5ccc(cc1n2)[nH]5)C=C4)[nH]3 and c1ccc2nsnc2c1 can form covalent organic frameworks through covalent bonding under certain conditions, which can utilize the excitation energy of singlet and triplet states for photocatalysis. Please determine which is the electron donor and which is the electron acceptor during the electron transfer process through the structural analysis of C1=Cc2cc3ccc(cc4nc(cc5ccc(cc1n2)[nH]5)C=C4)[nH]3 and c1ccc2nsnc2c1.
\end{itemize}
\paragraph{Analytical Chemistry:}
\begin{itemize}
    \item \citep{wu2025engineering} CC(C)(C)c1cc2cc(C(C)(C)C)cc3c4cc(C(C)(C)C)cc5cc(C(C)(C)C)cc(c(c1)c23)c54 is a fluorescent material. Please explain the reason why it can emit light from the perspective of molecular structure.
    \item \citep{guo2025acid} c1cc(-c2ccc3cc4cc(-c5ccncc5)ccc4cc3c2)ccn1 and N\#Cc1cc(C\#N)c(C\#N)cc1C\#N can be co-assembled into a eutectic and emit orange light under photoluminescence. After the addition of O=C(O)C(F)(F)F, the eutectic will undergo a transformation, and the luminescence will change from orange light to yellow light. Please explain the reason.
\end{itemize}
\paragraph{Polymer Chemistry:}
\begin{itemize}
    \item \citep{zhang2025manipulating} O=S(=O)(Oc1nc(=Cc2ccco2)c(OS(=O)(=O)C(F)(F)F)nc1=Cc1ccco1)C(F)(F)F and O=S(=O)(O c1nc(=Cc2ccccc2)c(OS(=O)(=O)C(F)(F)F)nc1=Cc1ccccc1)C(F)(F)F are monomers of two polymers that form polymers under solid-state heating conditions. In addition, it was found that O=S(=O)(Oc1nc (=Cc2ccco2)c(OS(=O)(=O)C(F)(F)F)nc1=Cc1ccco1)C(F)(F)F is relatively more prone to dimerization when heated in the solution. Please infer the reason by comparing their structures.
    \item \citep{rugh2025polysilaketals} Under certain conditions, C[Si]1(C)OCCOCCO1 can undergo anionic ring-opening polymerization to obtain polymers, which exhibit extremely high ionic conductivity in lithium battery electrolytes, surpassing poly (ethylene oxide). Please analyze the reasons from a structural perspective.
\end{itemize}


\section{More Examples of Reliable Human-AI Collaboration}\label{more_collab}

\begin{figure}
    \centering
    \includegraphics[width=0.9\linewidth]{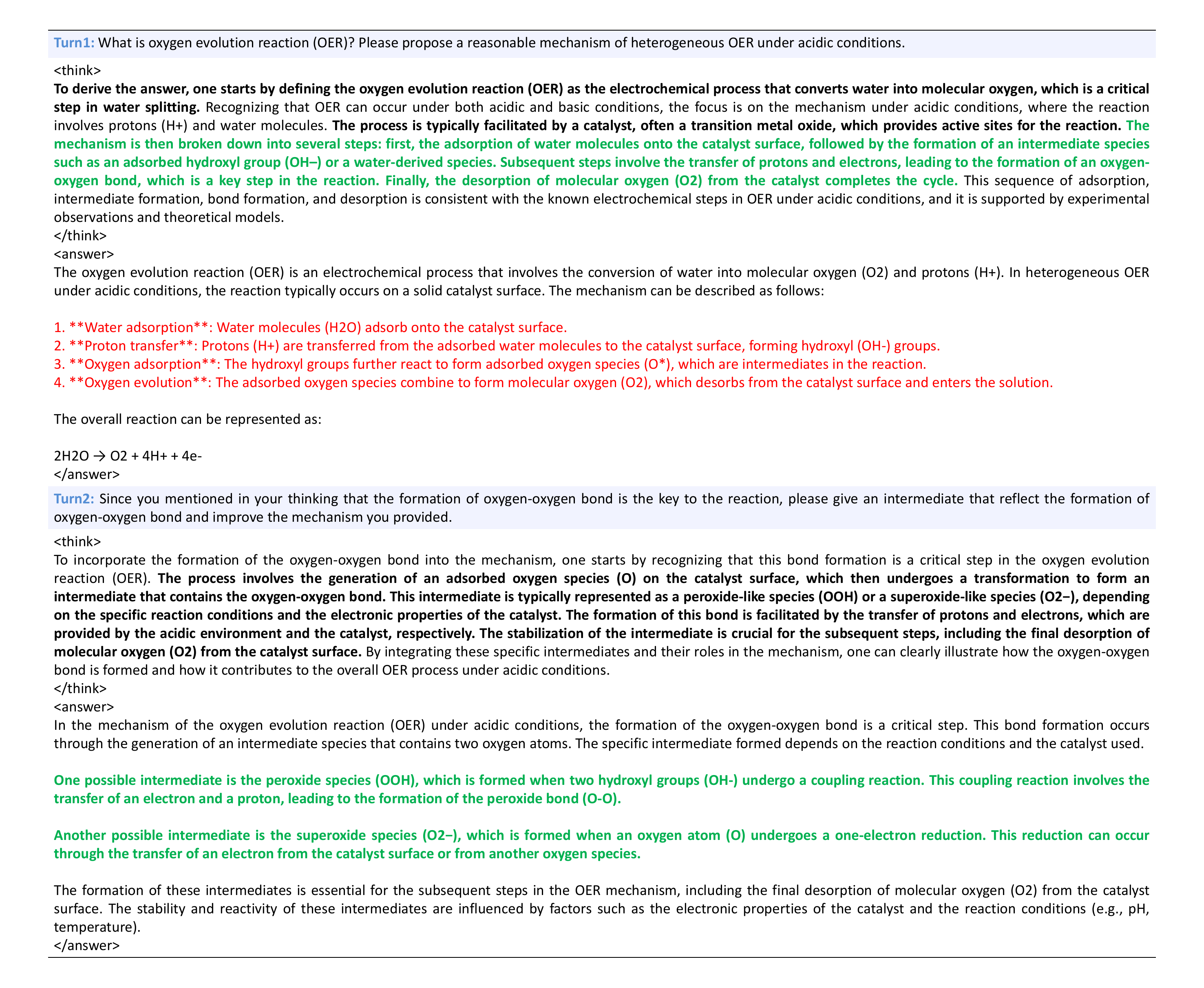} 
    \caption{An example of reliable human-AI collaboration using ChemDFM-R. We draw inspiration from this conversation from Li et. al. \citep{li2025achieving}. This example illustrates a process of error detection and correction with the help of rationales.}
    \label{fig:collab_1}
\end{figure}
\begin{figure}
    \centering
    \includegraphics[width=0.9\linewidth]{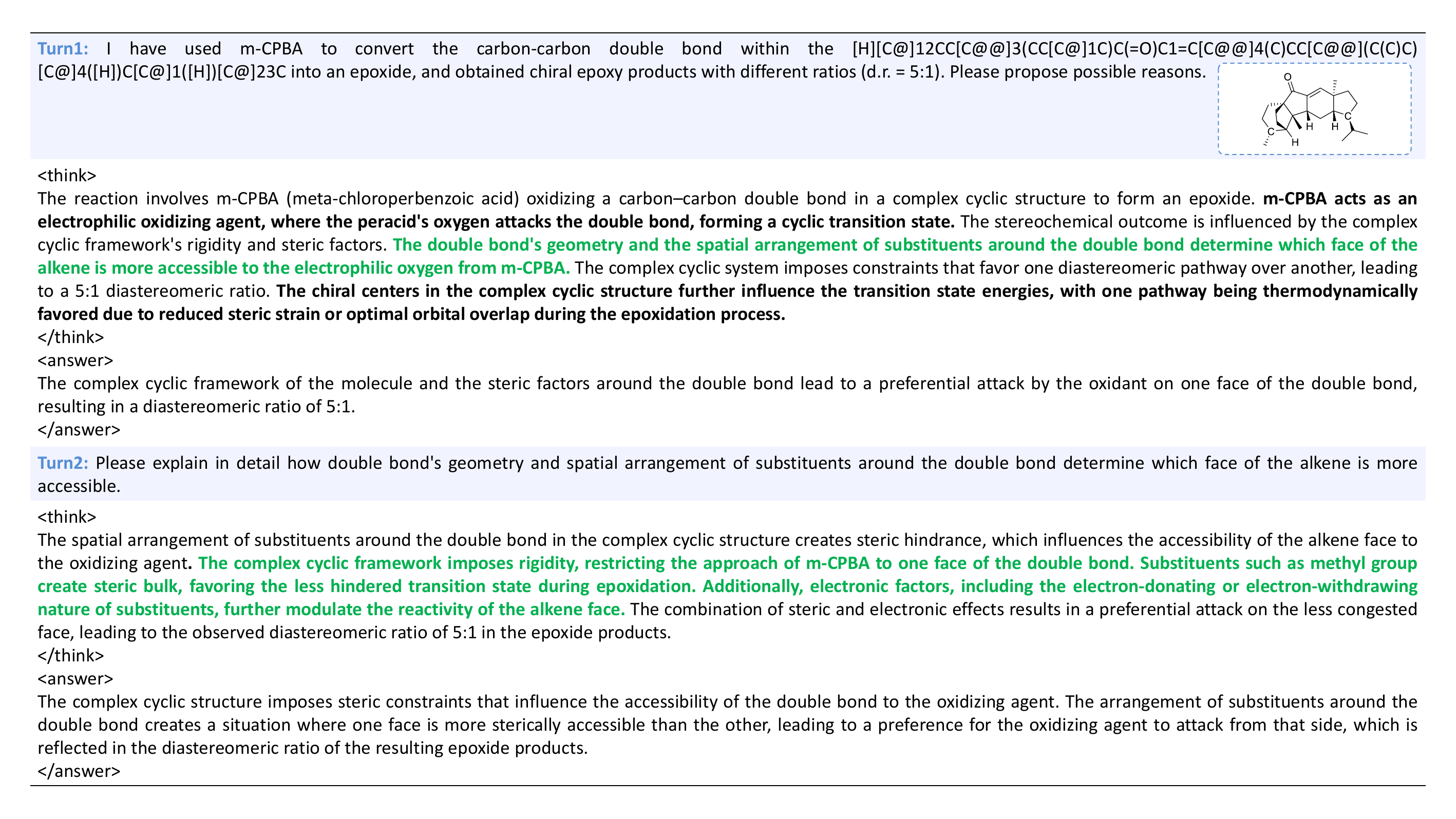} 
    \caption{An example of reliable human-AI collaboration using ChemDFM-R. We draw inspiration from this conversation from Yao et. al. \citep{yao2025convergent}. This example illustrates a process of information completion.}
    \label{fig:collab_2}
\end{figure}

Figure~\ref{fig:collab_1} illustrates a conversation starting from the same turn as that illustrated in the main text. In this conversation, we focus on fully understanding the mechanism of the oxygen evolution reaction~(OER). Suppose, as a newbie, we are unable to determine the correctness of the answer. With the help of ChemDFM-R's rationale, we could easily discover that the key step of the reaction mentioned in the rationale, which is ``the formation of an oxygen-oxygen bond'', is absent in the answer. This could serve as a reminder that the answer could be incorrect, and drive us to further request the model to clarify this inconsistency. After this follow-up inquiry, the model provided a better answer.

Figure~\ref{fig:collab_2} is a conversation about a reaction proposed in Yao et. al. \citep{yao2025convergent}. We first ask ChemDFM-R to explain the reason for the different ratios of the chiral epoxy products. Although the model's answer is relatively broad, ChemDFM-R thoroughly analyzes the influence factors in its rationale, including the bond's geometry and the spatial arrangement of substituents around the double bond, which is not included in the answer. With this information, we can further pursue the follow-up question and obtain an improved answer.